\documentclass[%
 reprint,
 amsmath,amssymb,
 aps,
 pra,
 floatfix,
 nofootinbib,
]{revtex4-2}

\usepackage{graphicx}
\usepackage{dcolumn}
\usepackage{bm}
\usepackage{gensymb}
\usepackage[]{tabularx} 
\usepackage[]{xcolor} 
\usepackage{hyperref}
\hypersetup{
    colorlinks=true,
    linkcolor=blue,
    filecolor=magenta,  
    citecolor=blue,    
    urlcolor=blue,
} 
\usepackage{cancel}
\usepackage{orcidlink}

\mathchardef\mhyphen="2D   

\newcommand{\York}{Department of Physics and Astronomy, York University, 4700 Keele St. Toronto ON, Canada M3J 1P3}
\newcommand{\JQI}{Joint Quantum Institute, Department of Physics, University of Maryland, College Park, Maryland 20742, USA}
\newcommand{\NIST}{National Institute of Standards and Technology, Gaithersburg, Maryland 20899, USA}

\newcommand{\vsig}{\langle v\sigma_{\rm spin}\rangle}
\DeclareMathOperator{\erf}{erf}
\DeclareMathOperator{\erfi}{erfi}
\DeclareMathOperator{\real}{Re}
\DeclareMathOperator{\im}{Im}

\begin{document}

\preprint{APS/123-QED}

\title{ Measurements of diffusion coefficients for rubidium--inert gas mixtures using coherent scattering from optically pumped population gratings}

\author{Alexander Pouliot\,\orcidlink{0000-0003-2569-5358}}
\email{alexpouliot@live.com}
 \affiliation{\York}
\author{Eduardo Chomen Ramos\,\orcidlink{0009-0001-0040-8300}}%
\affiliation{\York}
\author{Gehrig Carlse\,\orcidlink{0000-0003-4423-8478}}%
\affiliation{\York}
\author{Thomas Vacheresse\,\orcidlink{0009-0001-3352-3006}}%
\affiliation{\York}
\author{Jaskaran Randhawa\,\orcidlink{0009-0009-6880-9950}}%
\affiliation{\York}
\author{Louis Marmet\,\orcidlink{0000-0001-6087-5480}}%
\affiliation{\York}
\author{A. Kumarakrishnan\,\orcidlink{0000-0002-6874-1361}}%
\email{akumar@york.ca}
\affiliation{\York}
\author{Jacek K{\l}os\,\orcidlink{0000-0002-7407-303X}}
\affiliation{\JQI}
\author{Eite Tiesinga\,\orcidlink{0000-0003-0192-5585}}
\email{eite.tiesinga@nist.gov}
\affiliation{\NIST}\affiliation{\JQI}

\date{January 23, 2025}

\begin{abstract}
We present comprehensive determinations of the diffusion coefficients $D$ at $T=24\,\degree$C for trace amounts of naturally abundant Rb atoms in inert, naturally abundant He, Ne, N$_2$, Ar, Kr, and Xe buffer gases 
using a single measurement technique.
They have been measured by establishing a spatially periodic population grating in the Rb sample using two laser beams that intersect at a small angle $\theta$  of a few milliradians.
The atomic population grating decays exponentially in time due to diffusive motion induced by
momentum-changing elastic collisions between Rb and buffer gas atoms or molecules. This decay is monitored by observing the scattered field from a read-out beam aligned along the direction of one of the excitation beams.
We distinguish the contribution of diffusion from other collisional processes by measuring the characteristic $\theta^2$ dependence of the decay rate. 
We also measure the systematic dependence of the decay rate on the buffer gas pressure over a  range of $7\,000$ Pa to $90\,000$ Pa. 
In this manner, we obtain diffusion coefficients at standard atmospheric pressure of $101\,325$ Pa  and at a temperature of
24.0(5)~$^\circ$C.
We use two models to correct for systematic effects due to the transit time, one assuming a rectangular profile of the population distribution and a rectangular read-out beam profile, and a second using Gaussian profiles.
We obtain weighted averages of $0.33(5)$ cm$^2$/s, $0.214(14)$ cm$^2$/s,  $0.132(7)$ cm$^2$/s, $0.123(9)$ cm$^2$/s, $0.093(9)$ cm$^2$/s, and $0.073(4)$ cm$^2$/s for Rb in He, Ne, N$_2$, Ar, Kr, and Xe, respectively. 
The number in parentheses represents one standard deviation of the combined statistical and systematic uncertainty.
We also compare this data with diffusion coefficients obtained  using quantum, 
classical, and semi-classical theoretical methods based on the most accurate interatomic interaction potentials from the literature.
Near room temperature, simulations of $D$ using classical and quantum methods agree within their intrinsic, sub-1\,\% standard uncertainties.
We find that the semi-classical model only gives the correct orders of magnitude for $D$.
Our computed diffusion coefficients based on the quantum theory agree with the experimental determinations when systematic effects are taken into account.
Our measurements and modeling are relevant to the optimization of magnetometers, biomedical imaging using spin-polarized noble gases, tests of collision models based on interatomic potentials, and the development of pressure sensors.
\end{abstract}

\maketitle

\section{\label{sec:intro}Introduction}

Compact atomic vapor cells containing trace amounts of alkali-metal atoms in mixtures of inert buffer gases have become essential platforms for cutting-edge quantum sensors and precision metrology. 
While their utility has long been recognized for Spin-Exchange Relaxation Free (SERF) 
magnetometry~\cite{Allred}, a technique that has evolved into the most precise method for measuring magnetic fields,
the use of spin-polarized noble gases for medical imaging~\cite{HapperPolImg},  schemes for quantum memory~\cite{Lvovsky2009} and  vapor cell atomic clocks~\cite{AbdelHafiz2017} have increased their utility. 
Vapor cells are also used to study optical pumping and coherent transient effects~\cite{McNeal,Franz,Happer,Wagshul,EricksonThesis,Parniak2014,Mossberg1979,Forber1983}.

The collision-induced broadening of spectroscopic parameters, binary collision cross-sections, and binary diffusion coefficients for alkali-metal atoms in inert buffer gasses are 
key parameters for determining the many-body evolution in these vapor cell 
systems~\cite{Chapman}.
Because these observables are ultimately determined by the electronic potential energy 
surfaces between alkali-metal atoms and inert buffer-gas atoms or molecules, 
\cite{BlankWeeksKedziora2012,Medvedev2018,Klos,Xantheas2025}
measurements may be compared to {\it ab initio} calculations.
A limited sampling of attempts to reconcile calculations with measurements of cross-sections and spectroscopic parameters using atomic beams can be found in Refs.~\cite{RabiCrossSection,Rothe1959}.  
Similar comparisons in vapor cells and cold atomic samples can be found in Refs.~\cite{Miller2016,Mlynek,Gibble1991}   and Refs.~\cite{SadeghpourSpin2009,SawyerYe2008,DereviankoBabb2010,Gibble2013}, respectively.
A different approach involves measurements and predictions for spectroscopic broadening and shift parameters due to collisions.~\cite{Belov1981,CollBroad,RomalisBroad,Myneni2023}.

In this paper, we have exploited and broadened the development of a recently demonstrated coherent transient technique~\cite{Diffusion} to measure the binary diffusion coefficients near room temperature for trace amounts of rubidium in the inert gases helium, neon, argon, krypton, xenon, and molecular nitrogen.
Our measurements are performed in naturally abundant rubidium vapor at $24.0(5)\,\degree$C with buffer gas pressures $p$ between 50 Torr and 700 Torr, where $1\ {\rm Torr}=133.322$ Pa \cite{siunit}.

\begin{figure}
	\centering
	\includegraphics[width=\linewidth]{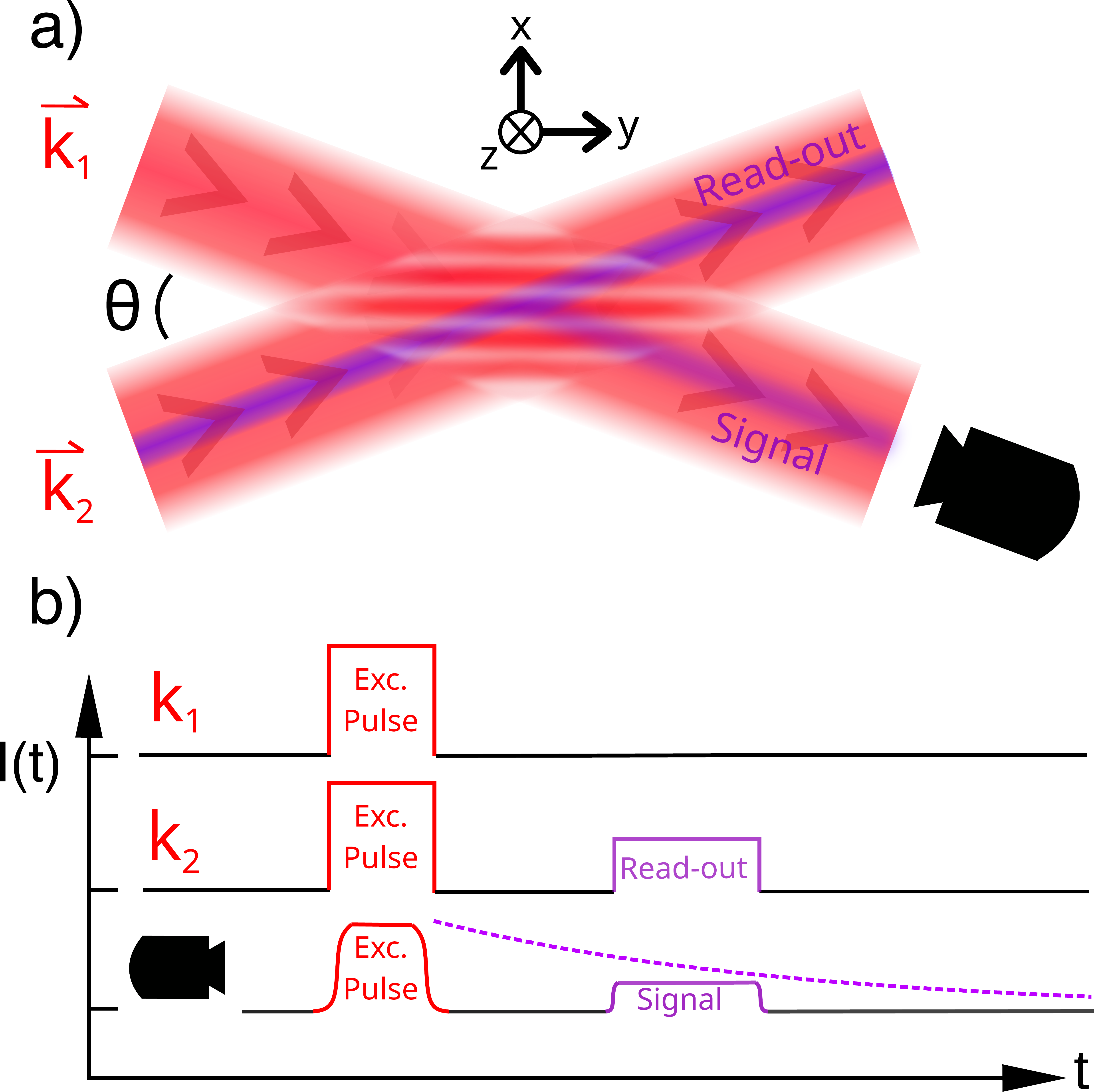}
	\caption { (color online)
 a) Diagram of the optical polarization grating  formed by two perpendicular linear polarized laser beams  with wavevectors $\vec k_1$ and $\vec k_2$ intersecting at a small angle $\theta$. One laser is polarized in the plane, the other is polarized out of plane.  The polarization grating forms along $\vec k_1 - \vec k_2$  and is indicated by red and white stripes.
 An excitation pulse is applied along both directions creating a spatial population grating in each $m$-level of the Rb sample along the same direction.
 A read-out pulse is applied along $\vec{k_2}$ with variable time delay after the excitation pulse, inducing a coherent scattered signal from the grating in direction $\vec{k_1}$. Fields applied at the time of the excitation pulse are shown in red, while those present at the time of the read-out pulse are shown in purple. All beams have the same spatial profile, but the read-out and signal beams are shown narrower so that they do not obscure the excitation beams.
 b) Timing diagram for the experiment showing the laser pulses applied along directions $\vec k_1$ and $\vec k_2$ as well as the intensity of the detected fields. The detector records light incident along  direction $\vec k_1$.
 The integrated signal pulse at each read-out delay is indicated by the dotted purple line.
     }
	\label{fig:grating}
\end{figure}

Our experiment is schematically shown in Fig.~\ref{fig:grating} as established in Ref.~\cite{Diffusion,myThesis}. 
An optical field with spatially periodic polarization is created by overlapping two laser pulses with perpendicular linear polarizations that intersect at a small angle $\theta$. 
The laser producing both of these excitation beams is locked 60~MHz below the frequency resonant on the $F=3 \rightarrow F'=4$ hyperfine transition of the $5s(^2{\rm S}_{1/2}) \rightarrow 5p(^2{\rm P}_{3/2})$ or D2 line  in $^{85}$Rb atoms in vacuum.
The excitation beam directions are aligned along $\vec k_1$ and $\vec k_2$, respectively. Their wavelengths $\lambda$ and wavenumbers $k=2\pi/\lambda$ are identical. 
This beam geometry produces a spatially periodic polarization grating along the direction $\vec k_1 - \vec k_2$ with a period of $\approx\lambda/\theta$.
The angle can be varied between 1.5 mrad and 4.0 mrad so that the number of periods ranges between six and fifteen across the 3~mm spatial extent of the beams in this direction.
Due to optical pumping in Rb, spatially periodic gratings are formed in the populations of the magnetic sublevels $m_F$ of the electronic ground state~\cite{Diffusion}.
Here, the quantization axis for the $m_F$ sublevels is along the lattice direction $\vec k_1-\vec k_2$.
These Rb population gratings have the same period $\approx\lambda/\theta$, and direction as the polarization grating. 

When a read-out pulse is applied later along the direction $\vec k_2$, phase-matched coherent scattering from the atomic lattice results in a scattered light signal along $\vec k_1$~\cite{bermanLaserPhys}.

The decay of this signal is exponential and has contributions from diffusion and from spin-exchange or spin-destruction collisions, all of which arise from the effect of elastic momentum-changing collisions of Rb with the buffer gas atoms or molecules. 
Other weaker effects, such as decoherence due to residual light, can also contribute to the decay.
Our technique, however, is able to differentiate between diffusion and other decay mechanisms 
by varying  angle $\theta$.
Since the grating period scales as $\theta^{-1}$, diffusion leads to a decay time constant $\tau$ that is proportional to $\theta^{-2}$.
We measure decay rates $1/\tau$ with a precision of a few percent and correct for the effects of the transit time.
Unlike our previous measurements, which were carried out at a fixed pressure~\cite{Diffusion}, we perform these experiments in an apparatus that can be filled with a buffer gas to a pressure of up to one atmosphere. 

We interpret our data on decay rates using the diffusion equation applied to the evolution of number density gradients~\cite{bermanLaserPhys}.
We compare the extracted diffusion coefficients with our theoretical calculations of thermalized diffusion coefficients expressed in terms of the microscopic differential cross section~\cite{Marrero1972} determined using accurate inter-atomic and inter-molecular potentials.
These comparisons reveal a systematic offset between the theoretical and experimental results.
A better understanding of this discrepancy can serve as the basis for an accurate pressure sensor with an operating range between 50 Torr to 1000 Torr.
Our studies complement efforts to realize pressure standards in ultra-high vacuum (UHV) environments using
alkali-metal atoms~\cite{Turlapov2016,Tiesinga2017,Shen2020,Comparison2022}. In these efforts either lithium 
or rubidium atoms are laser cooled to temperatures of tens of microkelvin.
\\

The diffusion coefficient is a function of both temperature $T$ and pressure $p$ and is given by
\begin{equation}
 D(T,p)=\frac{{\cal Q}(T)}{n}={\cal Q}(T)\frac{k_{\rm B}T}{p}.
 \label{eq:definition}
\end{equation} 
Accordingly, we report the diffusion coefficient at standard atmospheric pressure $p_0\equiv101\,325$~Pa
for the buffer gas, following the convention in the field of pressure metrology.
Here, quantity ${\cal Q}(T)$ is the pressure invariant diffusion coefficient, and is solely a function of $T$, $n$ is the number density of the inert gas, and we have used the ideal gas law $p=nk_{\rm B}T$ for the second equality in Eq.~\ref{eq:definition}. 
Another common representation of Eq.~\ref{eq:definition} in the literature is $D(T,p)=D_{\rm 0}p_{\rm 0}/p$, where  $D_{\rm 0}$ is the temperature-dependent diffusion coefficient at pressure $p_{\rm 0}$.

In Table~\ref{tab:compare}, we  summarize our experimental results on diffusion coefficients $D(T,p)$ 
for trace amounts of naturally abundant Rb interacting with buffer gasses $X={\rm He}$, Ne, N$_2$, Ar, Kr, and  Xe. 
In Table~\ref{tab:compare}, we also compare our measurements to existing experimental determinations relying on different physical effects, such as optical pumping, spin echoes, and coherent scattering. 
The values in Table~\ref{tab:compare} have been scaled to our measurement temperature of $T=24.0^\circ$C using a buffer-gas dependent power law obtained from our theoretical calculations.
Our values are among the most precise and confirm our previous measurement for nitrogen N$_2$ \cite{Diffusion} obtained at a fixed pressure. 
The data in this paper are obtained with a single experimental technique, with minimal changes required to interchange buffer gases, resulting in consistency of potential systematic errors across the six systems.
This allows for more complete comparisons with our theoretical results, and can serve as a benchmark for future work.

\begin{table*}
	\centering
	\caption{
 Comparison of measurements of diffusion coefficient $D(T,p)$ in cm$^2$/s at standard atmospheric pressure of $101\,325$ Pa for several naturally abundant Rb-inert gas mixtures. The second column gives $D(T,p)$  at $T=24.0(5)\,\degree$C scaled using the original data and temperature given in the third and fourth columns. We use $D\propto T^\beta/p$, where $\beta= 1.793$, 1.756, 1.730, 1.731, 1.750, and 1.791  for He, Ne, N$_2$, Ar, Kr, and Xe, respectively, as extracted from our theoretical models and explained in the text.
     Standard statistical uncertainties are given in parenthesis where available.
     We quote two results from this work for each gas, the first being the result without systematic corrections applied, and showing the purely statistical uncertainty in parentheses.
     The second and our recommended result is a weighted average of two different models for the transit time correction. 
     The first of these models uses rectangular spatial profiles for the population distribution and the readout pulse and the second model uses Gaussian profiles.
     The error in the corrected value includes estimates of systematic errors due to the transit time correction and wavefront curvature.
}
	\label{tab:compare}
	\renewcommand{\arraystretch}{1.3} 
	\begin{tabular}{|p{0.33\linewidth}p{0.2\linewidth}p{0.2\linewidth}p{0.2\linewidth}|}
		\hline
		\hline
		Buffer gas & \(D ~ (@\, 24\,\degree\textrm{C})\) & \(D\) (Orig. Data) & Orig. Temperature \\ 
		\hline
		\hline
		Helium (this work) & 0.33(2);\(0.33(5)\) & \(0.33(2)\);0.33(5) & $24 \,\degree$C\\
		Helium (Franz {\it et al.} 1976)~\cite{Franz1976} & 0.40 &  \(0.42\) & \(32\, \degree\)C\\
		Helium (Aymar {\it et al.} 1969)~\cite{Aymar69} & 0.41 &  \(0.41\) & \(27\, \degree\)C\\
		Helium (Franz 1965)~\cite{Franz65} & 0.53 &  \(0.68\) & \(67\,\degree\)C\\
		Helium (Bernheim 1961)~\cite{Bernheim62} & 0.46 &  \(0.54\) & \(50\,\degree\)C\\
		\hline
		Neon (this work) & 0.213(7);\(0.214(14)\) &  0.213(7);\(0.214(14)\) & $24\,\degree{\rm C}$\\
		Neon (Parniak {\it et al.} 2014)~\cite{Parniak2014} & \(0.202(5)\)   & \(0.222(5)\) & \(40\,\degree\)C \\
		Neon (Chrapkiewicz {\it et al.} 2014)~\cite{Chrapkiewicz2014} & \(0.19(2)\)   & \(0.24(3)\)& \(70\,\degree\)C \\
		Neon (Shuker {\it et al.} 2008)~\cite{Shuker2008} & 0.11  & \(0.13\) & \(52\,\degree\)C \\
		Neon (Franz {\it et al.} 1976)~\cite{Franz1976} & 0.22  & \(0.235\) & \(32\,\degree\)C \\
		Neon (Franz 1965)\cite{Franz65} & 0.38  & \(0.48\) & \(67\,\degree\)C \\
		Neon (Franzen 1959)~\cite{Franzen1959} & 0.27  & \(0.31\) & \(47\,\degree\)C \\
		\hline
		Nitrogen (this work) & 0.131(3);\(0.132(7)\) &  0.131(3);\(0.132(7)\) & $24\,\degree{\rm C}$ \\
		Nitrogen (Pouliot {\it et al.} 2021)~\cite{Diffusion}&0.129(1); $0.129(4) $ & 0.1490(14);\(0.149(5)\) & $50\,\degree$C \\
		Nitrogen (Ishikawa {\it et al.} 2000)~\cite{Ishikawa} & \(0.1305(16)\)  &\(0.159(4)\)& \(60\,\degree\)C \\
		Nitrogen (Erickson 2000)~\cite{EricksonThesis} &\(0.1446(19)\)   & \(0.30(3)\)& \(180\,\degree\)C \\
		Nitrogen (Wagshul {\it et al.} 1994)~\cite{Wagshul} &  0.15  &0.28 & \(150\,\degree\)C\\
		Nitrogen (Zeng {\it et al.} 1985)~\cite{Happer} & 0.16   &0.20 &\(70\,\degree\)C \\
		Nitrogen (Franz {\it et al.} 1976)~\cite{Franz1976} & 0.15  & 0.16 & \(32\,\degree\)C \\
		\hline
		Argon (this work) & 0.122(5)\(0.123(9)\) & 0.122(5)\(0.123(9)\) & $24 \,\degree {\rm C}$ \\ 
		Argon (Franz {\it et al.} 1976)~\cite{Franz1976} & 0.15 & \(0.16\) & \(32\,\degree\)C \\
		Argon (Bouchiat {\it et al.} 1972)~\cite{Bouchiat1972} & \(0.137(11)\)  & \(0.139(11)\) & \(27\,\degree\)C \\
		Argon (Franz 1965)~\cite{Franz65} & 0.29 & \(0.37\) & \(67\,\degree\)C \\
		Argon (Franzen 1959)~\cite{Franzen1959} & 0.21 & \(0.21\) & \(47\,\degree\)C \\
		\hline
		Krypton (this work) & 0.092(5);\(0.093(9)\)  & 0.092(5);\(0.093(9)\) & $24\,\degree {\rm C}$\\
		Krypton (Parniak {\it et al.} 2014)~\cite{Parniak2014} & \(0.085(3)\)   & \(0.093(3)\) & \(40\,\degree\)C \\
		Krypton (Chrapkiewicz {\it et al.} 2014)~\cite{Chrapkiewicz2014}& \(0.069(5)\)   & \(0.089(6)\) & \(70\,\degree\)C \\
		Krypton (Higginbottom {\it et al.} 2012)~\cite{Higginbottom2012} & \(0.033(5)\)  & \(0.043(7)\) & \(70\,\degree\)C \\
		Krypton (Bouchiat {\it et al.} 1972)~\cite{Bouchiat1972} & \(0.118(10)\)  & \(0.12(1)\) & \(27\,\degree\)C \\
		\hline
		Xenon (this work) & 0.072(3);\(0.073(4)\)  & 0.072(3);\(0.073(4)\) & $24\,\degree{\rm C}$\\
		Xenon (Parniak {\it et al.} 2014)~\cite{Parniak2014} & \(0.0519(9)\)    & \(0.057(1)\) & \(40\,\degree\)C  \\
		Xenon (Chrapkiewicz {\it et al.} 2014)~\cite{Chrapkiewicz2014} & \(0.053(3)\)   & \(0.068(4)\) & \(70\,\degree\)C \\
		\hline
	\end{tabular}
\end{table*}

The remainder of the paper is organized as follows.
In Sec.~\ref{sec:theory} we describe the theoretical basis for the formation and decay of the amplitudes of the atomic population gratings in the presence of a buffer gas. 
We  describe in Sec.~\ref{sec:model} how $D(T,p)$ is determined by microscopic scattering phase shifts and the differential cross section as derived within the Chapman-Enskog formalism \cite{Chapman,Hamel1986}.
In Sec.~\ref{sec:theoryresults} we present the predictions of quantum, semi-classical, and classical models of the differential cross sections for  noble gas atoms and N$_2$. We also compare the theoretical predictions of Rb-N$_2$ diffusion coefficients, with experimental determinations over the past 50 years.
In Sec.~\ref{sec:Experiment} we describe the details of the apparatus, including the vacuum and gas delivery system, the angle determination method, and the heterodyne detection of the coherently scattered light signal used to obtain our experimental results.
In Sec.~\ref{sec:results}, we present the experimental results, compare with theory, discuss systematic effects and potential applications such as the realization of a pressure standard. 
We conclude with a discussion of the impact of these results as well as their limitations in Sec~\ref{sec:conc}.

\section{\label{sec:theory}Formation and decay of optically pumped gratings}

\subsection{Simulations of grating formation}

For a quantization axis along the lattice direction ${\vec k_1} - {\vec k_2}$, the polarization formed by the small angle geometry in Fig.~\ref{fig:grating} is primarily linear everywhere, but has a small component which oscillates between $\sigma^+$ and $\sigma^+-$ polarizations, with a period of $\approx \lambda/\theta$.

We have simulated the spatially periodic optical pumping for this small angle geometry using rate equations~\cite{SPIEDefense,BermanRateEq} for the $F$ and $m_F$ populations of $^{85}$Rb and $^{87}$Rb in N$_2$ buffer gas incorporating collisionally broadened and shifted atomic resonances~\cite{CollBroad} as well as relaxation from collisions between an electronically excited Rb atom and buffer gas atoms or molecules~\cite{Speller1979}.
These parameters are the same within experimental uncertainties for all $F$ and $m_F$ levels and for both Rb isotopes.
References~\cite{Belov1981} and~\cite{CollBroad} measure these values for the buffer gases used in this work at a temperature of $T=20\,\degree$C and $T=121\,\degree$C, respectively. 
Although these rates are generally similar, they do not typically agree within their error bars.
For nitrogen, the broadening parameters are in agreement, and the shift parameters differ by only two standard deviations.
Here, we have used $18.9(5)$ MHz/Torr and $8.2(6)$ MHz/Torr for  pressure broadening and shifts, respectively, from Ref.~\cite{Belov1981} because the temperature in that experiment is better matched to the temperature at which our experiments are carried out. 
For the entirety of our pressure range, collisional broadening is on the order of several GHz, which is larger than the natural linewidth of Rb ($\approx6$ MHz), Doppler broadening ($\approx500$ MHz), and hyperfine splittings of the $5p(^2{\rm P}_{3/2})$
state ($\approx 300$ MHz) for either isotope. If the pressure exceeds $\approx250$ Torr, collisional broadening
is also larger than the $5s(^2{\rm S}_{1/2})$ hyperfine splittings ($\approx5$ GHz ).
The simulations represent a simplified model of the experiment as they, for example, ignore effects of the transit of atoms in and out of the volume of the excitation light and ignore coherences between magnetic sublevels.
Nevertheless, the simulations provide useful insight into  the population gratings in our Rb-inert gas mixtures.

The results of these simulations are shown in Fig.~\ref{fig:mLevels} for typical laser parameters  used in our experiments, and a representative absorption spectrum is shown in Sec.~\ref{sec:Experiment}.  
For these parameters, we find that the populations reach steady-state values so that a larger power or pulse length will not change the distributions.
We plot the deviation in $m_F$ level populations from a uniform distribution, $N_F/(2F + 1)$ for each level, for projections $m_F$ defined along $\vec k_1-\vec k_2$.
The simulations reveal that the spatial gratings of the magnetic sublevels have different amplitudes due to asymmetries in the coupling strengths of the excitation fields. 
The sum of all $F$ and $m_F$ populations adds to one at all positions corresponding to a constant Rb density profile.
The simulations show that the contrast in the $F=2$ ground state of $^{85}$Rb is significantly larger than that for $F=3$.
Similarly, the contrast for the $F=1$  state of $^{87}$Rb is larger than that for $F=2$.
Although the contrast of each $F, m_F$ grating is small, it is still possible to detect a signal because of the phase-matched emission of coherently scattered light,  whose intensity exhibits a faster than linear scaling with grating contrast~\cite{bermanLaserPhys}.

\subsection{Formation of the population gratings in collision-free and collisionally broadened regimes}

\begin{figure*}
	\centering
	\includegraphics[width=0.9\linewidth]{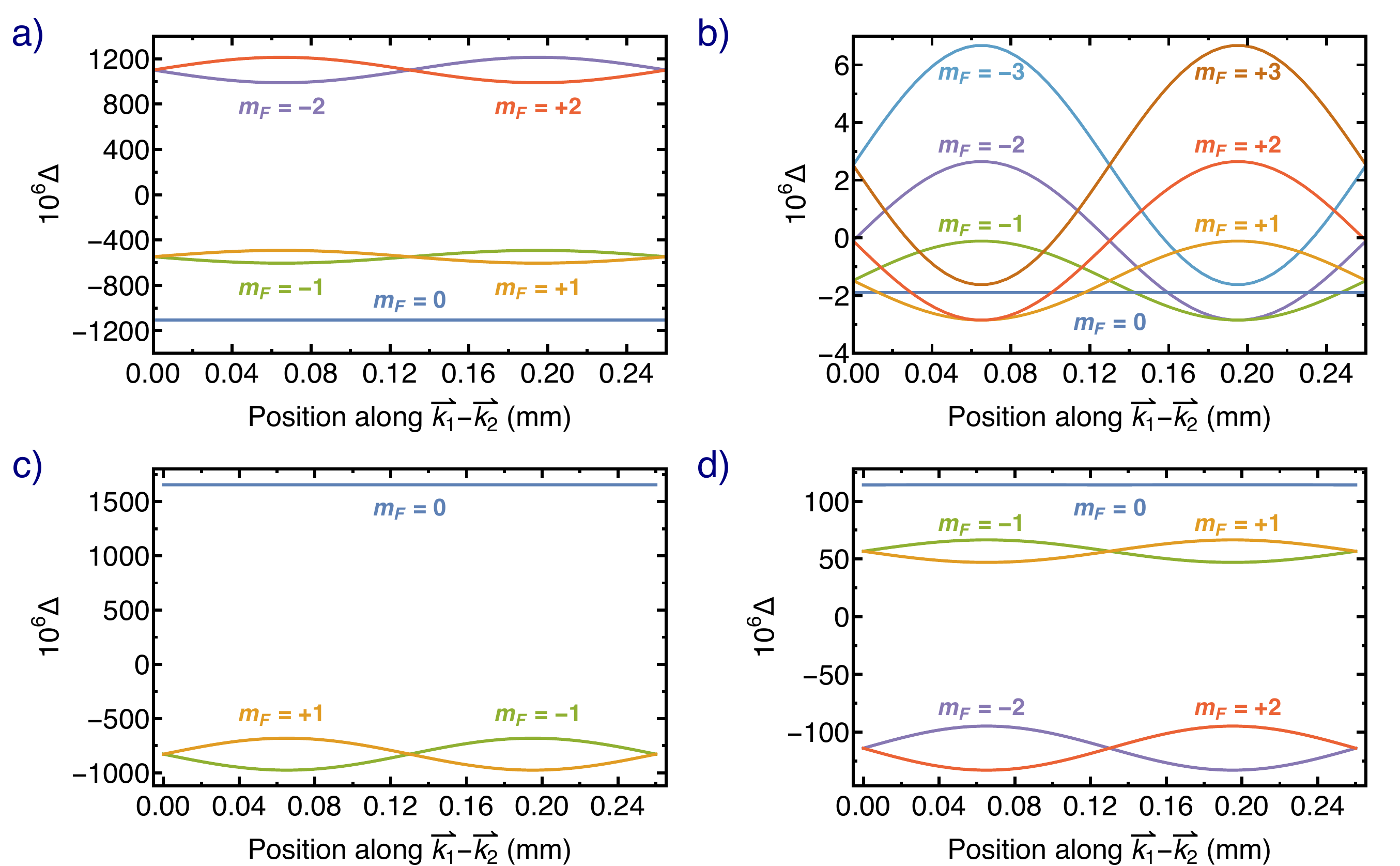}
	\caption{(Color online)  Simulation of the  population distribution after a 50 $\mu$s excitation pulse in  the ground-state magnetic sublevels quantized along ${\vec k_1}-{\vec k_2}$ of $^{85}$Rb $F = 2$ (panel a), $F = 3$ (b) and of $^{87}$Rb $F = 1$ (c), $F = 2$ (d) as functions of position along $\vec k_1-\vec k_2$ for one period of the grating. We have assumed a 200 Torr N$_2$ buffer gas and an average light intensity of 120 mW/cm$^2$. 
 The distribution for each level is shown in terms of its deviation $\Delta$ from a uniform population in each 
  $m_F$ sublevel for each $F$ state (i.e. a population of 1/(2$F$+1)).
 After the pulse, 96\,\% of the $^{85}$Rb population is pumped into the $F=2$ ground state and 4\,\% into the $F = 3$ state.
 For $^{87}$Rb the percentages are 87\,\% and 13\,\% for $F = 1$ and 2, respectively.
	For all four panels, the two linearly polarized laser beams intersect at angle $\theta=3$ mrad and have a  laser frequency that coincides with the $F=3 \rightarrow F'=4$ transition for a $^{85}$Rb atom in vacuum. 
}
	\label{fig:mLevels}
\end{figure*}

Although all of the experiments described in this work are performed in the presence of a buffer gas, it is useful to describe the formation of Rb population gratings  in both the presence and absence of buffer gas collisions for the small angle excitation geometry shown in Fig.~\ref{fig:grating}.
In the absence of a buffer gas and a low Rb vapor pressure of $3.5 \times 10^{-7}$ Torr at $24\,\degree$C, the Rb spatial population distribution can be assumed to evolve ballistically over the period of our gratings.
Then the grating amplitude in a room temperature Rb-only vapor cell, decays on a timescale of $1$ $\mu$s when $\theta\approx 1$ mrad~\cite{Diffusion}  due to the thermal motion of the Rb atoms, also known as Doppler dephasing.
This timescale  is significantly shorter than the ${\approx 10}$ $\mu$s transit time for room-temperature Rb atoms to pass through our 3 mm diameter laser beams. 
All velocity classes along ${\vec k_1}- \vec k_2$ are resonant with the excitation pulses, whereas short optical pulses can ensure that a substantial fraction of the Doppler broadened velocity distribution is excited along the direction of the excitation beams.
Although pulses as short as 100 ns are required, the optical pumping process remains incomplete and the grating contrast is significantly smaller than that shown in Fig~\ref{fig:mLevels}. 
Under these conditions, scattered light signals from coherence gratings, which have the same period as population gratings and are related to superpositions of adjacent $m_F$ levels, are dominant~\cite{bermanLaserPhys}.  
These signals have been exploited to measure velocity distributions, magnetic fields, and collision cross-sections in vapor cells, laser cooled gases, and atomic beams~\cite{NYUVapour,NYUTrap,Iain2008,NYUBeam}.

When a still small concentration of buffer gas corresponding to a pressure of $\approx10^{-5}$~Torr is added to the Rb vapor cell, the mean-free path for Rb-$X$ collisions  becomes of the order of the spatial period of the population grating. This buffer gas pressure  marks the beginning of the diffusive regime~\cite{BermanPRA94}.
When the buffer gas pressure is further increased to $\approx100$ Torr, the rubidium resonances are collisionally broadened and shifted, such that the excitation and read-out laser pulses couple to all ground and excited hyperfine states of both $^{85}$Rb and $^{87}$Rb. 
The transit time of $\approx 2$ ms at these pressures is then significantly larger than in the collision-free regime. 
If the excitation pulses are sufficiently long in the tens of $\mu$s regime, population gratings like those shown in Fig.~\ref{fig:mLevels} are produced. 
Further, when ambient magnetic fields  are minimized, the reflectivity of the grating does not change as a result of ground-state coherences.
The on-resonance optical depth of the sample, however,
is reduced by a factor of $\approx300$ compared to that of a Rb vapor cell without a buffer gas at the same temperature.
This occurs because the coupling strengths of the Rb hyperfine transitions are distributed over a large spectral width~\cite{Corney}, resulting in a reduction in the scattered light signal.
We compensate by increasing the Rb number density by about a factor of 10 by heating the cell.

\subsection{Decay of population gratings in diffusive regime}
In the diffusive regime, the ground-state populations $\rho_{Fm_F}(\vec x,t)\in[0,1]$ 
for  hyperfine state $F,m_F$ of $^{85}$Rb or $^{87}$Rb
at position $\vec x=(x,y,z)$ and time $t$ satisfy the diffusion equation
\begin{equation}
\frac{\partial\rho_{Fm_F} (\vec x,t)}{\partial t}=-D(T,p)\nabla^2\rho_{Fm_F}(\vec x,t)
\label{eq:diffusion}
\end{equation}
with coordinates $x$, $y$, and $z$ along directions $\vec k_1-\vec k_2$, $(\vec k_1+\vec k_2)/2$, and $\vec k_1\times \vec k_2$, respectively, as shown in Fig.~\ref{fig:grating}, and initial spatial dependence 
\begin{eqnarray}
\label{eq:population}
\rho_{Fm_F}(\vec x,t=0) &=& A_{Fm_F}+B_{Fm_F}\sin( k \theta x ) 
\end{eqnarray}
where  $0<|B_{Fm_F}|\ll A_{Fm_F}$ as suggested by our simulations in Fig.~\ref{fig:mLevels}.
Here,  the diffusion coefficient $D(T,p)$ is independent of the hyperfine state of either Rb isotope as we will show in Sec.~\ref{sec:model}.

When the read-out pulse is applied along $\vec k_2$, as shown in Fig.~\ref{fig:grating}, the phase matched emission from dipole oscillators interferes constructively along $\vec k_1$~\cite{bermanLaserPhys}. 
The amplitude of the electric field of the scattered light into this latter direction is measured and is proportional 
 to the $\vec k_1-\vec k_2$  Fourier component of the populations $\rho_{Fm_F} (\vec x,t)$. 
This signal is
\begin{equation}
E_{\rm C}(t)=E_0 e^{-(k\theta)^2D(T,p)t}
\label{eq:diffsig}
\end{equation}
with diffusion time constant $\tau_{\rm D}\equiv D(T,p)^{-1}(k\theta)^{-2}$
and initial non-negative amplitude $E_0$ which depends on the grating contrast ($B_{F,m_F}$ in Eq.~\ref{eq:population}) and the readout field $E_{\rm RO}$.
The corresponding decay rate $1/\tau_D$ scales inversely as a function of the buffer gas pressure as shown by Eq.~\ref{eq:definition}.

In our experiments, we record the decay rate of optically pumped population gratings as a function of the angle $\theta$ and the pressure of the buffer gas.
There are, however, two additional contributions to the decay rate. The first is due to binary spin-exchange or spin-destruction collisions between Rb and the inert gas atoms or molecules~\cite{HapperSpin,Saam92,Walker2001}.
This process adds a decay rate $n\vsig=p\vsig/(k_{\rm B}T)$ to the total decay of the grating.
Here, $\vsig$ is the thermally averaged rate coefficient for spin-exchange or spin-destruction collisions.

The second additional contribution is caused by residual laser light after the excitation beams are nominally turned off. Measurements have shown that the intensity of this residual
light is no larger than 0.1 mW/cm$^2$, less than 0.3 \% of the typical intensity of the excitation pulse. 
If the residual intensity along the directions $\vec k_1$  and $\vec k_2$ are closely balanced, the light will contribute to optical pumping and increase the amplitude of the Rb population grating. For unbalanced residual intensities, we find that the light will reduce the grating amplitude. 
These effects can be modelled by an additional contribution to the decay rate  given 
by  ${\cal W}(\nu,T,p) \Gamma_{\rm opt}$, where ${\cal W}(\nu,T,p)$ is a collisionally broadened and shifted Voigt profile and $\Gamma_{\rm opt}$ is a signed optical pumping rate, a function of both residual intensities~\cite{bermanTextbook}.

The dimensionless Voigt profile describes the atomic response to the laser light as a function of the laser frequency $\nu$ at fixed temperature and  buffer gas pressure $p$~\cite{Corney}. This normalized function is a convolution of a Gaussian distribution,  primarily due to Doppler broadening, and a Lorentzian distribution with a width that is primarily due to collisional broadening and thus proportional to $p$. For our temperature, and range of pressures, the Lorentzian component is dominant, and therefore ${\cal W}(\nu,T,p)$ has a $1/p$ dependence near atomic resonance. Therefore this term can  be approximated as ${\cal W}(\nu,T,p)\approx{\cal W}(\nu,T,p_{\rm ref})\times (p_{\rm ref}/p)$. Here, $p_{\rm ref}$ is a convenient reference pressure.

The resulting total decay rate of the population grating is 
\begin{equation}
\label{eq:generaldecay}
	\frac{1}{\tau}=\frac{p}{k_{\rm B}T} \vsig + D(T,p) (k\theta)^2 + {\cal W} (\nu,T,p)\Gamma_{\rm opt}
 \end{equation}
This expression can be rewritten to explicitly show the dependence on pressure by using Eq.~\ref{eq:definition}, and assuming that ${\cal W}(\nu,T,p)\propto 1/p$. Hence,
 \begin{equation}
	\label{eq:fulldecay}	
    \frac{1}{\tau}=\frac{\vsig}{k_{\rm B}T} p + {\cal Q}(T)k_{\rm B}T \frac{(k\theta)^2}{p} + {\cal W} (\nu,T,p_{\rm ref})\Gamma_{\rm opt}\frac{p_{\rm ref}}{p} \,.
\end{equation}

The diffusion coefficient can be unambiguously inferred from Eq.~\ref{eq:fulldecay} by measuring  this characteristic angle dependence of $\tau$. 
For decreasing  angle $\theta$, however, the number of grating planes in the excitation volume decreases,  optical pumping becomes less efficient and the angle determination is more error prone.
For increasing $\theta$, the decay time becomes shorter, which limits the precision.
Therefore, we find a balance by operating the experiment over the range $\theta=1.5$ mrad to 4 mrad.
In addition, $D(T,p)$ can also be measured by varying the pressure of the buffer gas when the contributions from spin-exchange, spin-destruction processes, and residual light are small.

\section{\label{sec:model}Models for calculating diffusion coefficients and differential cross sections}

We have performed numerical and approximate analytical simulations of differential cross sections as functions of collision angles and collision energy and of thermalized diffusion coefficients $D(T,p)$ of trace amounts of rubidium in a denser gas of inert atoms or molecules in the viscous pressure regime, derived within the Chapman-Enskog formalism \cite{Chapman,Hamel1986}.
We have determined $D(T,p)$ as functions of temperature 
$T$ assuming that the components of the gas mixtures have the same temperature.

The calculations for $^{85}{\rm Rb}+X$ as well as $^{87}{\rm Rb}+X$ are based on the electronic potential energy surfaces published  by Ref.~\cite{Klos} as well as those by Ref.~\cite{Medvedev2018}. The relevant potentials have well depths $D_{\rm e}$ expressed in energy equivalent temperatures $D_{\rm e}/k_{\rm B}= 1.6$ K, 8.8 K, 34 K, 56 K, 97 K, and 150 K
for $X={^4{\rm He}}$, $^{20}$Ne, $^{14}$N$_2$, $^{40}$Ar, $^{84}$Kr,
and $^{132}$Xe, respectively, and are shallow compared to those of more typical chemical bonds.
Moreover, these well depths are smaller than $k_{\rm B}T$ for the temperatures $T$ used in experiments measuring diffusion coefficients in the viscous regime. 
The standard uncertainties in $D_{\rm e}$ and, in fact, of the shape of the electronic potential energy surfaces can be found in Ref.~\cite{Klos}.

The electronic potential energy surfaces are isotropic for Rb  colliding with the spin-less noble-gas atoms. That is, the potentials only depend on the separation between 
the center of masses of the atoms. The mechanical relative orbital angular momentum is then conserved and 
coupling between partial waves is absent. Isotropic interaction potentials imply that differential cross sections only depend on polar scattering angle $\Theta$. The potential energy surface of the three-atom system $^{87}$Rb-N$_2$ is anisotropic, but as shown in Ref.~\cite{Klos} the anisotropy is weak and leads to negligible transition rate coefficients between ro-vibrational states of N$_2$. These rate coefficients were found to be smaller than our standard uncertainties for the total rate coefficients. We can again assume that the mechanical relative orbital angular momentum is  conserved.
Finally, the non-zero electron and nuclear spins of ground-state $^{85}$Rb and $^{87}$Rb are bystanders for collisions near room temperature. In other words, hyperfine interactions can be ignored and $D(T,p)$ is independent of the total angular momentum $F$ and its projection $m_F$ of the Rb atom.

We can then use the expression for the diffusion coefficient from Ref.~\cite{Marrero1972}.
Assuming an isotropic, non-reactive, and spin-independent interaction, the diffusion coefficient is
given by Eq.~\ref{eq:definition}
with
\begin{equation}
{\cal Q}(T)=\frac{3}{16} \sqrt{\frac{2\pi k_{\rm B}T}{\mu}} \frac{1}{ {\cal A}(T)}
\label{eq:PInvariant}
\end{equation}
and reduced system mass $\mu$. 
The thermally averaged diffusion area ${\cal A}(T)$ is
\begin{eqnarray}
  {\cal A}(T) 
     &=& \frac{\displaystyle \int_0^\infty e^{-E/k_{\rm B}T} E^2 {\cal A}(E) {\rm d}E}{\displaystyle \int_0^\infty e^{-E/k_{\rm B}T} E^2  {\rm d}E} 
\end{eqnarray}
with the diffusion area $\mathcal{A}(E)$
\begin{eqnarray}
\label{eq:ACrossSec}
   {\cal A}(E) &=& \int {\rm d}\Omega  (1-\cos\Theta) \frac{{\rm d}\sigma}{{\rm d}\Omega} 
   \label{eq:AofE}\\
    &=&\frac{4\pi}{k_{\rm r}^2} \sum_{\ell=0}^\infty  (\ell+1) \sin^2 [\eta_{\ell+1}(E)-\eta_l(E)]\,,
\end{eqnarray}
where  $E=\hbar^2k_{\rm r}/(2\mu)$ is the collision energy with  collision wavenumber $k_{\rm r}$, ${\rm d}\sigma/{\rm d}\Omega$ is the differential elastic scattering cross section,
which now only depends on polar scattering angle $\Theta$, and
$\eta_\ell(E)$ is the collisional phase shift for partial wave $\ell$ and collision energy $E$.
The factor $1-\cos\Theta$ suppresses the role of small angles $\Theta$ in area ${\cal A}(T)$ as small angle collisions do not
lead to significant diffusion or transfer of momentum between Rb and inert gas atoms or molecules.

The theoretical simulations involve numerically calculating the regular solutions of radial Schr\"odinger
equations as functions of   
the separation $R$ between the center of masses of $^{87}$Rb and a noble gas atom or N$_2$. We have  one equation for each partial wave $\ell=0, 1, 2,\dots$. Collisional phase shifts $\eta_\ell(E)$ as functions of collision energy $E$ and $\ell$ are extracted from the large-$R$ behavior of these solutions \cite{Child}. The reduced mass $\mu$ in the kinetic energy operator of the Schr\"odinger
equations is determined from the atomic masses.
We have used the atomic masses from Ref.~\cite{AME2020} and recommended energy conversion factors 
from Ref.~\cite{CODATA2018}.
Standard uncertainties in the theoretical values for $D(T,p)$ are estimated from scattering calculations performed with the electronic potential energy surfaces obtained from {\em ab initio} calculations with several basis sets for the electronic wavefunctions as described in Ref.~\cite{Klos}.

In principle, we need to determine diffusion coefficients for all isotopologues of systems ${\rm Rb}+X$,
especially for those elements where more than one isotope has a significant abundance.
However, the electronic potential energy surfaces for different isotopologues
are the same to a good approximation and we need only to change the reduced mass in the
kinetic energy operator of the Schr\"odinger equations. The  modification in the potential is 
the so-called diagonal non-adiabatic correction~\cite{Lengsfield:1986}, which for our systems are smaller than the uncertainties in our potentials. The percent changes in the reduced mass appearing in the kinetic energy operator lead to small changes in the diffusion coefficients.
We verified this latter observation by performing calculations for 
$^{85}{\rm Rb} + {^{14}{\rm N}_2}$ and $^{87}{\rm Rb} + {^{14}{\rm N}_2}$. 
At $T=300$~K the diffusion coefficient is 0.3\,\% larger for $^{85}{\rm Rb} + {^{14}{\rm N}_2}$, which as we will show is 
of the same order of magnitude as the theoretical uncertainty in $D(T,p)$ due to the uncertainty in the potential energy surface. In fact, the small change in $D(T,p)$ for different isotopes is consistent with $D(T,p)\propto \mu^{-1/2}$ from an approximate analytical model described in the next section. 
Nevertheless, for the remainder of this paper we will only quote theoretical diffusion coefficients 
for $^{87}$Rb scattering from the most-abundant isotope of the relevant inert gas atom or molecule.

Finally, in the experimental data for Ar, Kr, and Xe buffer gasses, as also will be shown later on in this paper, we find evidence of a weak coupling between the spins of Rb and the rotation of the molecule similar to the observations by Ref.~\cite{Bouchiat1972}. 
That is, we find a finite but small value for rate coefficient $\vsig$.
These anisotropic couplings do not affect our theoretical results for $D(T,p)$ within our quoted uncertainties.

\section{Results of the simulations}\label{sec:theoryresults}

We have determined ${\cal A}(T)$ and thus $D(T,p)$ with {\em quantum}, {\em classical}, and {\em semi-classical} methods. The quantum calculations are divided into two parts. The first corresponds to  ``exact'' numerical solutions of Schr\"odinger equations using
the most-accurate electronic potentials from Refs.~\cite{Medvedev2018,Klos} for tens of thousands of pairs $\ell$ and $E$ with $\ell$ up to 950 for our heaviest system, and $E < k_{\rm B}\times 3000$ K.
The differences in values for $D(T,p)$ based on the potentials from Ref.~\cite{Medvedev2018} and those from Ref.~\cite{Klos} are smaller than our theoretical uncertainties.
For the second quantum results, we replace the shallow electronic potentials by ones that are purely repulsive. That is,
we keep the repulsive wall of the electronic potentials from Ref.~\cite{Klos} for ${V(R)> k_{\rm B}\times 45}$ K,
smoothly connected to a repulsive exponential potential for larger separations. We chose the potential
energy at the connection point to be smaller than our collision energies at room temperature.

The classical model is based on the expression for the classical deflection angle in terms of the impact parameter and the collision energy~\cite{Child,Medvedev2018}. The classical results are computed numerically using the potentials from Refs.~\cite{Medvedev2018,Klos}. 
Again, the differences in values for $D(T,p)$ based on the potentials from Ref.~\cite{Medvedev2018} and those from Ref.~\cite{Klos} are smaller than our theoretical uncertainties.
It is worth noting that within the classical approximation Monchick in Ref.~\cite{Monchick1959} found by inspection that for a repulsive exponential 
potential $V_{\rm exp}(R)=V_0\exp(-b R)$ for {\it all} $R$,  the thermally averaged diffusion coefficient is well described by 
\begin{equation}
	D(T,p) \propto \frac{(k_{\rm B}T)^{3/2}}{[\ln(V_0/k_{\rm B}T)]^2p}\,.
  \label{eq:repulsive}
\end{equation}
Here, $V_0$ and $b$ are the positive parameters of the exponential potential
(see also Section 2.3.b of Ref.~\cite{Marrero1972}). This scaling law is appropriate at temperatures above 300 K because the thermal kinetic energy is much larger than the depth of the attractive part of the interaction potentials.

Finally, the semi-classical method  is analytical, and solely relies on perturbative scattering from the attractive, long-range,  van-der-Waals $-C_6/R^6$  component of the electronic potential with a positive dispersion coefficient $C_6$. The methodology starts with  the Born approximation for the phase shift $\eta_\ell(E)= (3\pi/32) (E/E_6)^2/\ell^5$, which is valid for collision energies $E\gg E_6$ and  small polar angles ${\Theta\ll\pi}$ \cite{Child}.
Here, the van-der-Waals energy is $E_6=\hbar^2/(2\mu\beta_6^2)$ with van-der-Waals length $\beta_6=(2\mu C_6/\hbar^2)^{1/4}$,
 and $\hbar$ is the reduced Planck constant.
The van-der-Waals energies for our systems lie between $k_{\rm B}\times 0.14$ mK for $^{87}$Rb-Xe and $k_{\rm B}\times 29$ mK for $^{87}$Rb-He. With this Ansatz for $\eta_\ell(E)$ and changing the sum over $\ell$ into an integral, all relevant thermalized quantities can be computed analytically. Specifically, we have for the semiclassical thermalized diffusion area
\begin{equation}
	{\cal A_{\rm sc}}(T)=3.9728 (k_{\rm B}T/E_6)^{-1/3} \beta_6^2\,,
 \label{eq:AvdW}
\end{equation}
so that  $D(T,p)\propto (kT)^{5/6}\mu^{-1/2} C_6^{-1/3} p^{-1}$
(see also Section 2.3.a of Ref.~\cite{Marrero1972}).
Although approximate, this model sets expectations for the orders of magnitude of the diffusion coefficients.

\begin{figure}
\includegraphics[width=3.4in, trim=0 0 0 0,clip]{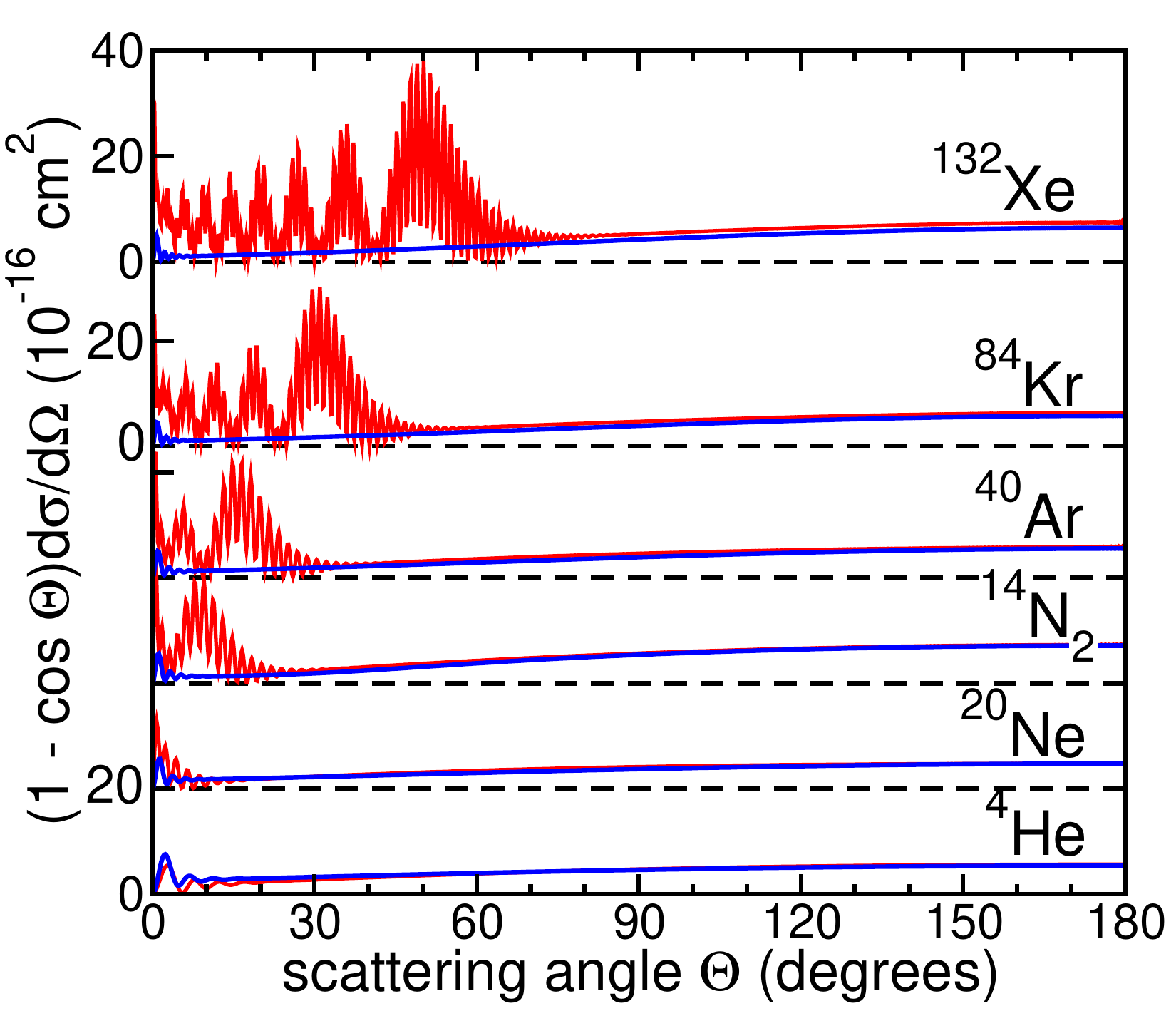}
\caption{(Color online)``Weighted'' differential cross sections $(1-\cos\Theta){\rm d}\sigma/{\rm d}\Omega$ at a collision energy of $k_{\rm B}\times 300$ K  as functions of polar scattering 
angle $\Theta$ for systems $^{87}{\rm Rb}+X$, where $X={^4{\rm He}}$, $^{20}$Ne, $^{14}$N$_2$, $^{40}$Ar, $^{84}$Kr,
and $^{132}$Xe. The red curves are obtained from  quantum scattering calculations
using the most-accurate electronic potentials by Ref.~\cite{Klos}.  
Blue curves are obtained with quantum scattering calculations using a purely repulsive potential derived from the repulsive wall of the potential for this system (See text for details).
Data for the various systems have the same vertical scale but are displaced by unequal amounts for clarity. Dashed black lines correspond to the zero values for the relevant data. 
}
\label{fig:dSdOmega}
\end{figure}

Figure \ref{fig:dSdOmega} shows  ``weighted''  differential cross sections $(1-\cos\Theta) {\rm d}\sigma/{\rm d}\Omega$ for our  $^{87}$Rb+$X$ systems as functions of polar scattering angle $\Theta$ for  collision energy ${E=k_{\rm B}\times 300}$ K. 
These differential cross sections are determined  using the quantum-mechanical method and using the most-accurate electronic potentials by Ref.~\cite{Klos}. 
They have a complex, fast oscillatory behavior for polar angles up to $15^\circ$ for $^{87}$Rb+He and even up to $60^\circ$ for $^{87}$Rb+Xe. Moreover, the amplitude of these oscillations increases with the mass of the inert gas atom or molecule. These oscillations correspond to glory scattering from interferences between waves reflected at  different atom-atom separations $R$ \cite{Child}. For larger angles the ``weighted'' differential cross sections are slowly increasing smooth functions. 
These behaviors are in sharp contrast to those for ${\rm d}\sigma/{\rm d}\Omega$, which is extremely large and highly peaked for $\Theta< 1^\circ$, {\it i.e.} near forward scattering.
Integrated over all angles, however, the total cross section $\sigma(E)=\int {\rm d}\Omega \,({\rm d}\sigma/{\rm d}\Omega)$ is only two to three times larger than the diffusion area ${\cal A}(E)$ at the same $E$. 
Unsurprisingly, ${\cal A}(E)< \sigma(E)$ as small angle $\Theta$ scattering does not
lead to significant diffusion or transfer of momentum between Rb and inert gas atoms or molecules.

Figure~\ref{fig:dSdOmega} also shows ${(1-\cos\Theta)} {\rm d}\sigma/{\rm d}\Omega$ for $^{87}{\rm Rb}+ X$ systems,
where the electronic potential has been replaced by the purely repulsive shape. We find that the rapid oscillations in the forward scattering region are mostly absent,  but that for larger polar scattering angles the ``weighted'' differential cross sections based on the ``exact'' potential and the repulsive potential are in agreement. 

\begin{figure}
\includegraphics[width=3.3in, trim=0 0 0 0,clip]{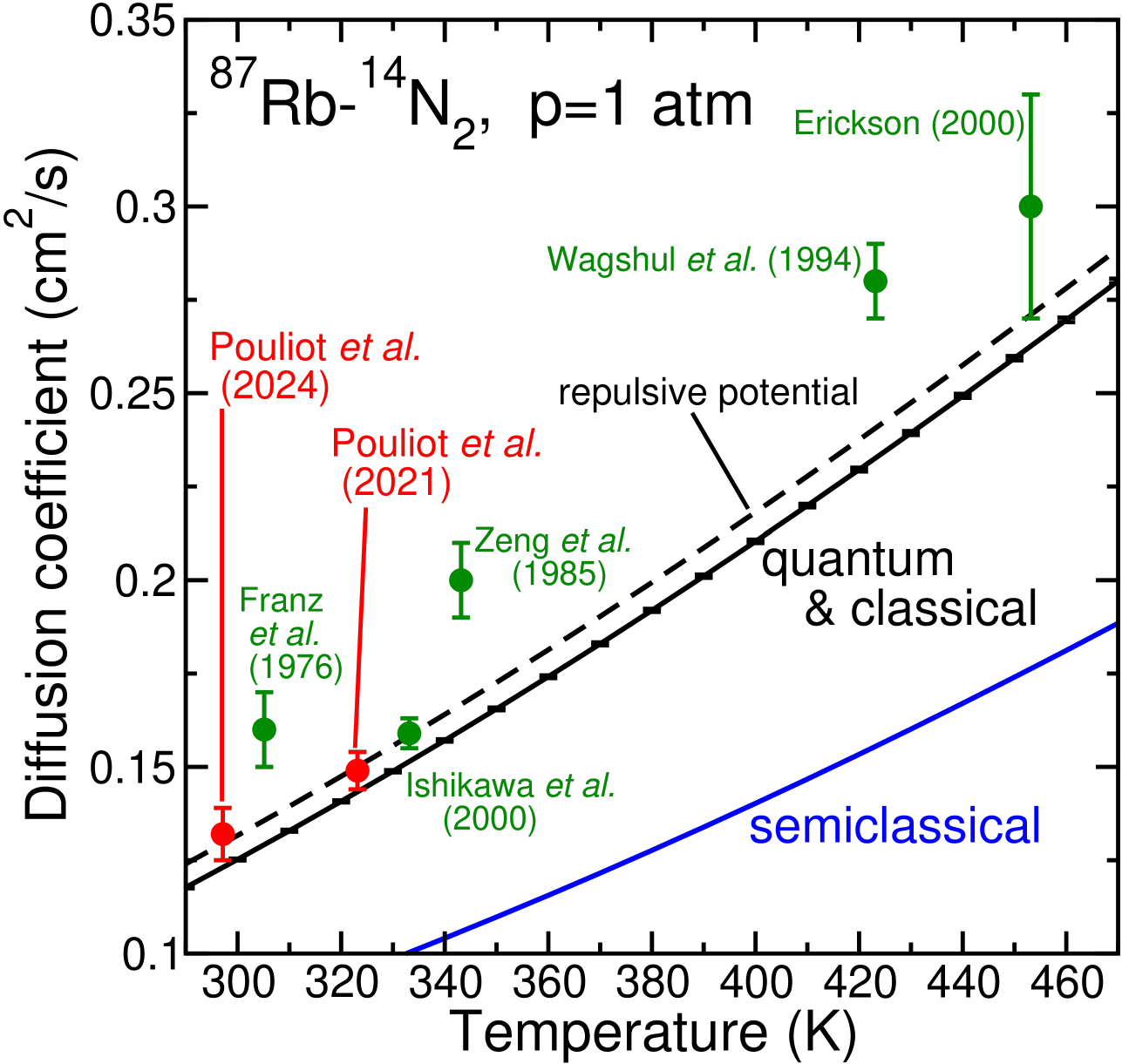}
\caption{(Color online) Thermalized diffusion coefficients  as functions of temperature $T$ for $^{87}$Rb+$^{14}$N$_2$
and at the standard atmospheric pressure of $p=101\,325$ Pa. The solid black curve with small standard-uncertainty error bars correspond to our quantum simulations as well as the (nearly) indistinguishable classical simulations from Ref.~\cite{Medvedev2018}.
The dashed black curve corresponds to a quantum simulation using the purely repulsive potential constructed from
our most-accurate electronic potential.
The blue curve corresponds to the semi-classical, analytical estimate. Two overlapping red markers with standard uncertainties just below $T=300$ K are  from Ref.~\cite{Diffusion} and our current measurements. 
Green markers  with standard uncertainties and author labels at higher temperatures correspond to measurements
found in Table~\ref{tab:compare}.
For publications that did not supply an uncertainty budget we assume standard uncertainty 1 in the last significant digit.
}
\label{fig:DiffRbN2}
\end{figure}

We compare thermalized diffusion coefficients $D(T,p)$ at standard atmospheric pressure as functions of temperature $T$ obtained with  our {\it quantum, classical}, and {\it semi-classical} theoretical methods , along with
those measured experimentally over the past fifty years for $^{87}{\rm Rb}+{^{14}{\rm N}}_2$ in Fig.~\ref{fig:DiffRbN2}.
Firstly, we observe that the quantum and classical diffusion coefficients based on the most-accurate potential energy surfaces are identical to within our uncertainties and are slowly increasing.
The less than 0.5\,\% standard uncertainties in $D(T,p)$ for the temperature range shown in the figure are due to uncertainties in the $^{87}$Rb-N$_2$ 
potential energy surface from Ref.~\cite{Klos}. 
For temperatures between 300 K and 500 K, classical simulations of $D(T,p)$ are more than adequate.  Now, however, we can quantify the degree of agreement using our quantum results.

Figure~\ref{fig:DiffRbN2} also shows $D(T,p)$ based on the quantum-mechanical calculations using  the purely repulsive potential.
These values are systematically larger than those based on calculations using the most-accurate potential energy surface but have visually identical temperature dependencies.
A larger $D(T,p)$ is consistent with the observations in Fig.~\ref{fig:dSdOmega} for  $^{87}{\rm Rb}+{^{14}{\rm N}}_2$.
That is, the difference in $D(T,p)$ for the two calculations are due to differences in the behavior of $(1-\cos\Theta) {\rm d}\sigma/{\rm d}\Omega$ at relatively small polar angles.

The observation regarding the temperature dependence of $D(T,p)$ in the previous paragraph leads us to try to fit the theoretical $^{87}{\rm Rb}+{^{14}{\rm N}}_2$ quantum data in Fig.~\ref{fig:DiffRbN2} to the analytical expression for $D(T,p)$ for an exponential potential given in Eq.~\ref{eq:repulsive}.
The fit with its two adjustable parameters reproduces our theoretical data to within their uncertainties, thereby validating the results of Ref.~\cite{Monchick1959}. A naive fit to $(k_{\rm B}T)^{3/2}$, however, is insufficient to reproduce our theoretical results.

Next, Fig.~\ref{fig:DiffRbN2}  shows the semi-classical, analytical estimates for $^{87}{\rm Rb}+{^{14}{\rm N}}_2$. They are significantly lower than the quantum and classical results, {\it e.g.}, by about 30\,\%, and have a noticeably weaker $T$ dependence than that found with the quantum or classical simulations. For the $^{87}$Rb and rare-gas-atom systems, the analytical estimates are  similarly too small. 
We conclude from the calculations with the purely repulsive potential and the semi-classical, analytical estimates
that the diffusion coefficients are mainly determined by the inner repulsive wall of the potential energy surfaces. 

Finally, Fig.~\ref{fig:DiffRbN2}  shows the available experimental diffusion coefficients for ${\rm Rb+N}_2$ as measured over the past fifty years and over a temperature range between 300 K and 460 K. Our current
measurement with this system is performed at 24 $^\circ$C and is the smallest studied temperature. The experimental data is consistently larger than the theoretical values. The  experimental data for other ${\rm Rb}+X$ systems do not necessarily follow this pattern, as we will show and discuss later on in this article.

Data for thermalized diffusion coefficients as functions of $T$ for all systems studied here, and over a larger temperature range, can be found in  Supplemental Information.

\section{\label{sec:Experiment}Experimental Details}

\begin{figure}
	\includegraphics[width= 0.9\linewidth]{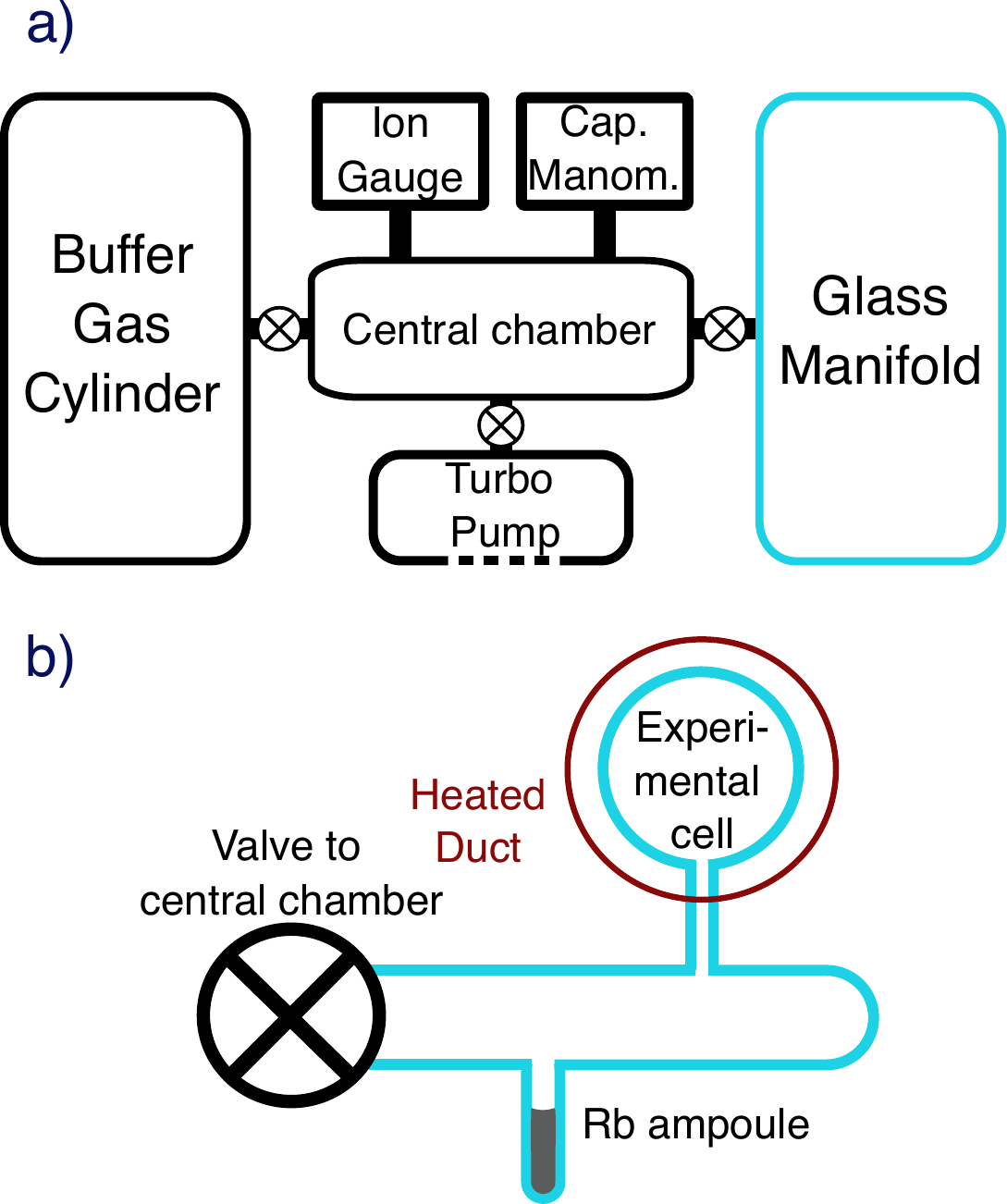}
	\caption{(Color online) a) Sketch of gas handling apparatus including the glass manifold. Gate valves are represented by circles with crosses. A capacitance manometer (``Cap. Manom.'' in the figure)  measures
 pressures in the central chamber between 0.1 Torr and 1000 Torr, while an ion gauge measures pressures from $10^{-9}$ Torr to $10^{-4}$ Torr.
 b) Detailed sketch of the glass manifold with the valve connecting the manifold to the central chamber shown in panel a). The cylindrical out-of-plane heated duct
 (red circle) helps maintain a constant rubidium vapor density in the experimental cell.
  }
	\label{fig:ManifDrawing}
\end{figure}

Our apparatus has two important components, namely the laser system and the vacuum system which includes the experimental cell containing the Rb-inert gas mixture.
The laser system consists of a home-built external-cavity diode laser (ECDL)~\cite{HerminaRSI} that seeds a tapered waveguide amplifier (TA) with $\approx$15~mW of power to realize an optical output of 2~W at the desired laser frequency \cite{SPIEPWest}. 
The ECDL is frequency stabilized with respect to a D2 line in $^{85}$Rb using saturated absorption spectroscopy in a 5 cm long Rb reference cell that contains no buffer gas.
The lock point coincides with the peak of the $F=3 \rightarrow F^\prime=3,4$ crossover resonance in $^{85}$Rb near room temperature, which is 60.3~MHz red detuned from the $F=3 \rightarrow F^\prime=4$ transition  for a $^{85}$Rb atom at rest.
The output of the TA is split into two beams, each is amplitude modulated by an 80-MHz acousto-optic modulator (AOM). 
The AOMs are driven by a radio frequency (RF) network consisting of an RF synthesizer, RF amplifiers, transistor-transistor logic (TTL)
switches, and pulse generators. 
The frequency up-shifted, 3 mm diameter laser beams emerging from the two AOMs are aligned along the nearly parallel directions $\vec k_1$ and $\vec k_2$, respectively, and passed through the atomic vapor cell.

The vacuum system is schematically shown in Fig.~\ref{fig:ManifDrawing}a with a more detailed diagram of the pyrex glass manifold shown in Fig.~\ref{fig:ManifDrawing}b.
The cylindrical experimental atomic vapor cell has a diameter of 5.1~cm, a length of 5.1~cm, and circular endfaces of thickness 0.3~cm.
This cell is attached to a pyrex manifold by a thin stem of length 4.4~cm and an outer diameter of 0.3~cm,
located halfway along the length of the cylinder.
The $\vec k_1$ and $\vec k_2$ laser beams are well collimated (with a Rayleigh range of $\approx 1$~m), aligned perpendicular to the endfaces of the experimental cell, and strongly overlapped over the cell length.
One end of the pyrex manifold is attached to a gate valve with a glass-to-metal graded seal.  
A glass ampoule containing a sample of natural isotopic abundance Rb is fused to the cylinder below the experimental cell as shown in Fig.~\ref{fig:ManifDrawing}b.
A gate valve attaches the glass manifold to a small stainless-steel central vacuum chamber. 
The glass manifold can be evacuated by a turbo pump and then filled with rubidium as described below.
The central chamber can also be filled with inert gases from a high-pressure gas cylinder via a stainless-steel gas line through a second gate valve. 

The pressure in the experimental cell is measured using a capacitance manometer in the central chamber with the connecting valve open. 
The manometer has an operating range between 0.1 Torr and 1000~Torr with at worst a 0.2\,\% fractional standard uncertainty. 
The central chamber and glass manifold can be evacuated to $\approx10^{-8}$ Torr by a turbo pump with a pumping speed of 70~L/s. Our base, or lowest-achievable pressure is $5 \times 10^{-9}$~Torr. 
When the vacuum system is operated below $10^{-4}$~Torr, {\it e.g.} while pumping out the central chamber and glass manifold, a micro-ion gauge is used to monitor the pressure. 
The turbo pump is attached to a roughing pump rated at 200~L/s through a fore-line bellows hose (not shown in Fig.~\ref{fig:ManifDrawing}). The fore-line pressure is monitored by a Pirani gauge.

Before a buffer gas can be introduced into the experimental cell from the central chamber, rubidium vapor is cycled into the experimental cell by repeatedly 
heating the rubidium ampoule and the glass walls of the manifold for ten minutes at a time, in order to create thermal gradients between the pyrex manifold and the experimental cell that speed up the introduction of the Rb vapor into the cell.
This loading procedure is necessary because the migration of rubidium is highly restricted due to collisions after the buffer-gas is added~\cite{HapperCoating}.

Stable and sufficiently large rubidium vapor pressures in the experimental cell are needed to optimize the observed signals.
This is accomplished by surrounding the cylindrical experimental cell with a heated cylindrical aluminum duct, 10.1~cm in diameter and protruding 20~cm on either side of the length of the experimental cell. 
The duct has a  circular hole of diameter 4~cm  to accommodate the thin glass stem of the experimental cell. 
The duct is wrapped in resistive heater tape and the air just inside the duct is maintained at a temperature of $50(2)\,\degree$C as measured by a thermocouple. 
As a result of the 4~cm hole in the duct, there is a steady-state convective  flow of air from the surrounding laboratory at $T_{\rm lab}=23$~$^\circ$C in through the hole and out through the ends of the heated duct.
Under these conditions, the walls of the experimental glass cell are coldest near the stem,
and hottest on the opposite, upper side. 

This configuration ensures that the rubidium density at the center of cell is enhanced by the vapor leaving the hottest surface.
The absorption spectra of a probe laser passing through the center of the cell indicates that the rubidium density is about a factor of 10 larger than its value without heating.
This aspect of the measurement will be discussed in more detail later in the paper.
Despite the increased Rb density, the gas mixture near the center of the cell illuminated by the laser beams is inferred to be at a temperature of $24.0(5)\,\degree$C.
This inference is supported by two separate considerations.
Firstly, we find that measurements of $D(T,p)$ in N$_2$ remain unchanged within experimental uncertainty of 1\,\% for a fixed angle and pressure, when the temperature of the duct is increased from 23$\,\degree$C to $90\,\degree$C.
In contrast, for this temperature variation, $D(T,p)$ is expected to increase by 42\,\% for N$_2$ based on the theoretical predictions and experimental data shown in Fig.~\ref{fig:DiffRbN2}.
A stable measured $D(T,p)$ with its 1\,\% accuracy then implies
that the temperature  of the gas mixture at the center of the cell can be no higher than $24\,\degree$C for the typical duct temperature of $50\,\degree$C. 
In addition, the room temperature of $23\,\degree$C places a lower bound on the gas temperature.
Secondly, a simulation based on the heat and the steady-state Navier-Stokes equations, that uses measured temperatures around the cell and the thermal conductivities of the gas and pyrex glass, provides confirmation of this temperature range.

We observe a gradual depletion of rubidium vapor in the experimental cell over the course of days due to 
the thermal gradients.
These observations are consistent with a non-equilibrium distribution of rubidium.
Under these conditions the rubidium density is strongly influenced by the hottest surfaces of the cell.
We monitor the rubidium density in the experimental cell by measuring collisionally broadened Rb spectra~\cite{Diffusion}.
When the Rb is depleted, we evacuate the system to base pressure and reintroduce rubidium and buffer gas.

We cancel ambient magnetic fields using three pairs of
square magnetic-field coils, oriented along  three orthogonal axes and with currents running in the same direction.
The coils in each pair are separated by 0.55$\ell$, where $\ell$ is the side length of the square, and are centered on the experimental cell. 
We cancel linear magnetic field gradients using three additional coil pairs in the same geometry.
These additional coils have currents in opposing directions.

As in previous work~\cite{Diffusion} a finite  magnetic field and gradient is initially set to easily record an oscillatory signal associated with Rb Larmor precession from a coherence grating magnetometer.
Subsequently, the field and gradient are iteratively reduced to zero to maximize the coherently scattered light signal from the Rb population grating. 
In the current work this iterative procedure is applied to all three axes, whereas in Ref.~\cite{Diffusion} we only reduced the magnetic field and gradient along the direction $\vec k_1-\vec k_2$. 

\begin{figure}
	\includegraphics[width=\linewidth]{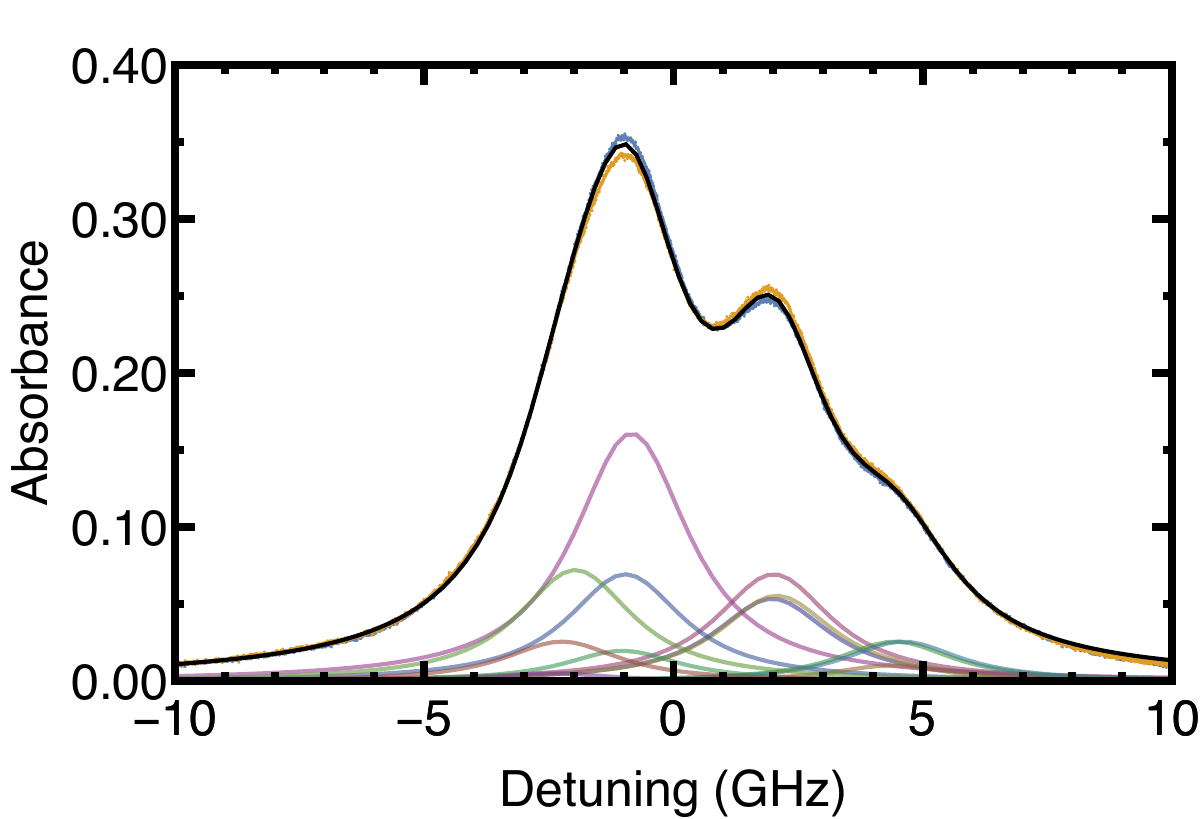}
	\caption{ Rubidium absorption spectra a cell of length 5~cm at $T=24.0(5)$ $^\circ$C and a pressure of 134.10(5) Torr of N$_2$ buffer gas as measured by the capacitance gauge. 
 The plot shows  absorbances $-\ln(I/I_0)$ as a function of laser detuning, where
   $I$ and $I_0$ are the transmitted and off-resonant transmitted laser intensities, respectively, and the laser detuning is relative to the $F=3 \rightarrow F'=4$ hyperfine transition frequency of the D2 line in $^{85}$Rb in its rest frame and  absent  any buffer gas.
		Blue and yellow traces correspond to data from upward and downward scans of the laser frequency. 
		A fit of both traces to the composite spectrum for collisionally broadened and shifted $^{85}$Rb and $^{87}$Rb D2 hyperfine transitions is shown in black~\cite{Diffusion}. 
	Individual D2 hyperfine line profiles  are also shown as colored lines, the most prominent of these colored lines is the collisionally broadened and shifted $F=3 \rightarrow F'=4$ line in $^{85}$Rb.
 }
	\label{fig:exSpec}
\end{figure}

The pressure of the inert gas in the experimental cell is measured using the capacitance manometer and is independently verified, typically to within 20\,\%, using atomic spectroscopy of Rb resonances~\cite{Diffusion}.  
In this latter approach, pressure is determined from a fit to the collisionally broadened and shifted rubidium lines.
Figure~\ref{fig:exSpec} shows an example of such a Rb spectrum for trace amounts of Rb in a N$_2$ buffer gas as well as a fit to the spectrum~\cite{Diffusion-Erratum}.
Using the shift and broadening parameters from Ref.~\cite{CollBroad}
and assuming that these parameters are independent of temperature for the relevant temperature range,
the fit gives a N$_2$ pressure of 145.7(5) Torr while the capacitance manometer reading is 134.10(5) Torr.
Moreover, we determine the Rb number density to be $ 8.2(4) \times 10^{10}$ cm$^{-3}$, from the fitted signal amplitude of the spectrum.
The Rb density can be inferred from the amplitude using the known length of the experimental cell, D2 transition dipole moments of $^{85}$Rb and $^{87}$Rb~\cite{Steck85,Steck87}, and the isotopic abundances of $^{85}$Rb and $^{87}$Rb.

\begin{figure}
    \centering
    \includegraphics[width=\linewidth]{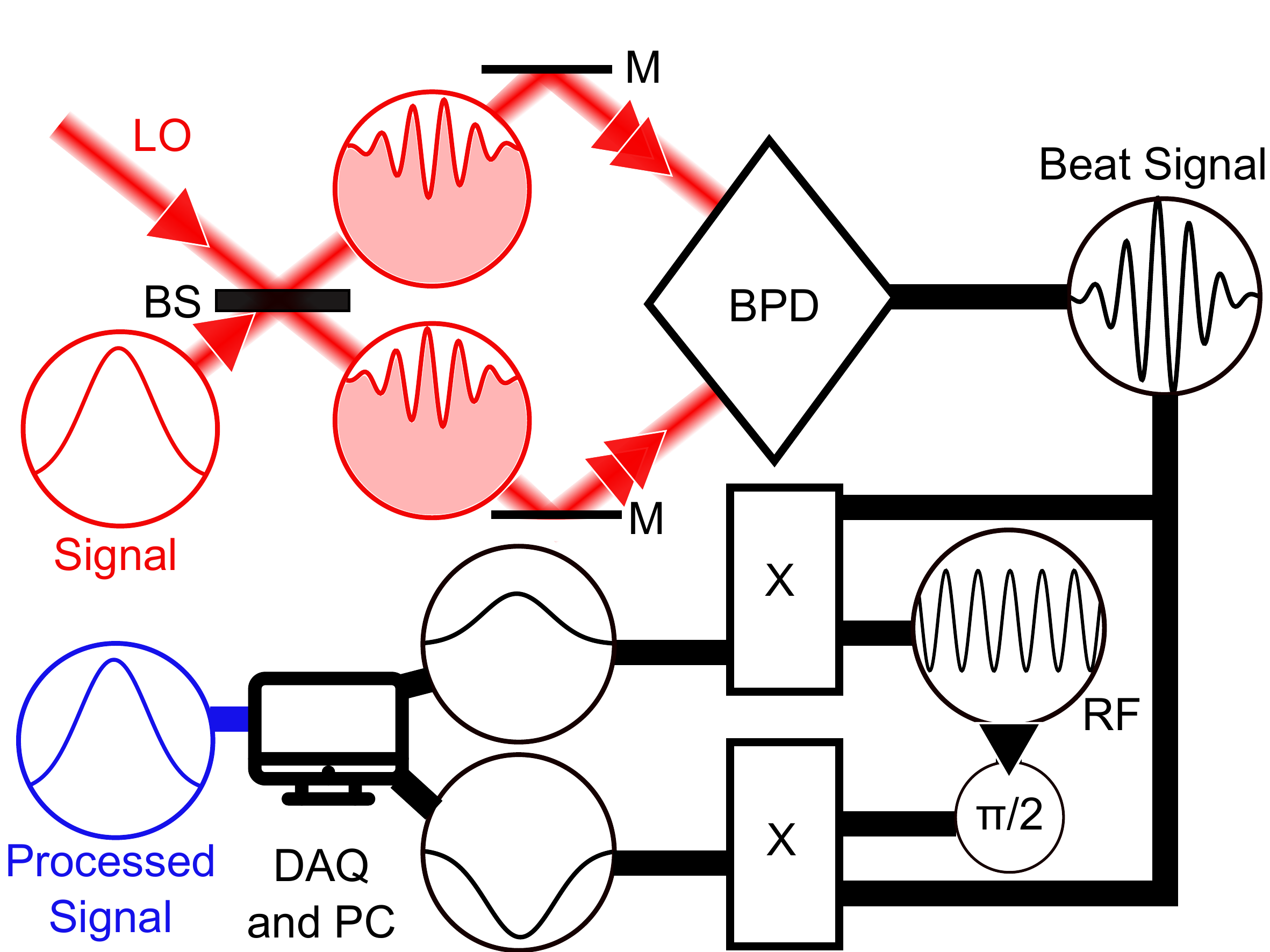}
    \caption{Simplified diagram of our heterodyne detection setup.
    Optical signals are shown in red, analog electronic signals are shown in black, and digital electronic signals are shown in blue. LO - local oscillator light from the ${\vec k_1}$ AOM;   BS - 50:50 beamsplitter; BPD - balanced photodiodes; M - mirrors; RF - RF signal used to drive the AOMs, and used here to demodulate the hetrodyne signal    $\pi/2$ - $\pi/2$ phase shifter; X - RF mixers.}
    \label{fig:heterodyne}
\end{figure}

The heterodyne detection technique is illustrated in Fig.~\ref{fig:heterodyne}. 
The undiffracted laser beam from the $\vec k_1$ AOM, at a frequency 80 MHz below the frequency of the diffracted $\vec k_1$ beam, serves as a local oscillator (LO) 
and travels through the air gap between the heated duct and the glass cell.
The LO is combined with the beam along $\vec k_1$ that passes through the center of the experimental cell on a beam-splitter downstream from the cell. 
Among other possible configurations, this path of the LO minimizes phase noise relative to the $\vec k_1$ beam. 
The two outputs of the beam splitter, which consist of heterodyne signals that have a relative $\pi$ phase shift and a beat frequency of 80 MHz, are incident on the two photodiodes of a balanced detector.
These Si:PIN photodiodes have 1~ns risetimes and are reverse-biased with voltages of opposite polarity.
As a result, the combined heterodyne signals of the two photodiodes add in phase, while their DC offsets cancel.
The combined 80 MHz signal from the photodiodes is gated by a TTL switch so as to shield the downstream amplifiers from the intense excitation pulse, while allowing the scattered signal to pass through at the time the read-out pulse is applied.
The signal from the switch is then amplified and mixed down to DC using the RF oscillator that drives the AOMs to generate the in-phase and $\pi/2$-out-of-phase components. 
The mixed down signal envelopes are further amplified and recorded using an 8-bit analog to digital converter (ADC) with a bandwidth of 125~MHz at a sample acquisition rate of $500\times10^6$ per second in each channel.
The details of the detection scheme are described in Fig.~4 of Ref.~\cite{Diffusion}.

The experiment is operated at an approximately 1 kHz repetition rate using digital delay generators. 
The time base of these generators synchronized to the 10~MHz clock of a radio-frequency synthesizer~\cite{SRSSynth,disclaimer}. 
The delay generators are also triggered by the clock signal to reduce jitter. 
The pulses from the delay generators are coupled to the AOMs using TTL switches with an RF extinction ratio of 80 dB. 
Nevertheless, this optical setup produces residual scattered light along  $\vec k_1 $ and $\vec k_2$ at the level of a few $\mu$W even when the AOMs are turned off.  
The level of the background light depends on alignment and varies on a timescale of days.

\begin{figure}
	\includegraphics[width=0.7\linewidth]{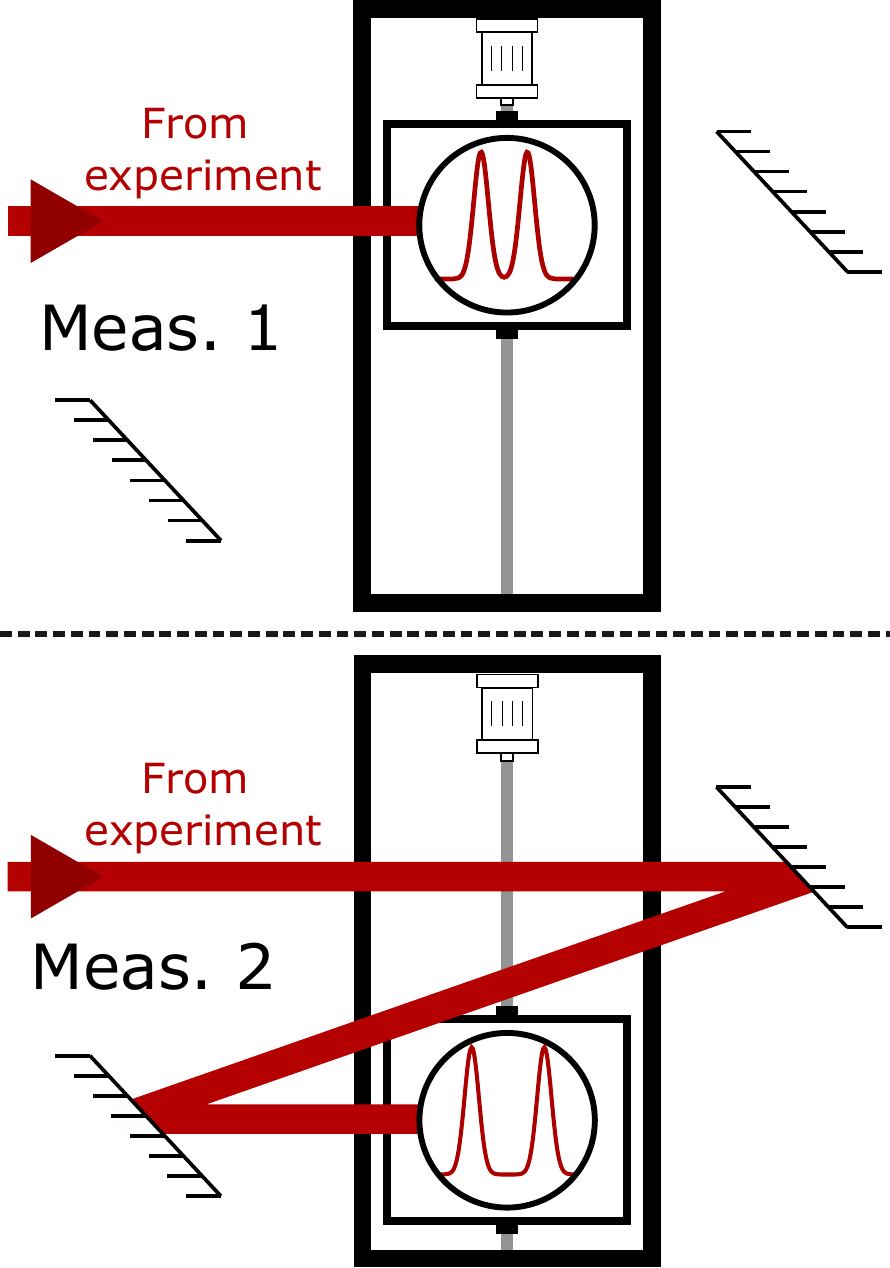}
	\caption{(Color online)
 Schematic representing the angle measurement protocol using two measurement steps. 
	The two nearly parallel lasers, indicated by red beam paths coming from the left, are operated in cw mode and a motorized translation stage (thick rectangular box) moves a spatial beam profiler (circle inside a square with red signal traces.) between two points along the beam path. 
The angle is calculated from Gaussian fits to the double-peaked beam-profiler signal. 
The  path distance between the two points is approximately 82 cm and is measured with a precision of $0.3$~cm.
}
	\label{fig:Profiler}
\end{figure}

The angle $\theta$ between $\vec k_1$ and $\vec k_2$ is measured using a beam profiler attached to an automated, motorized translation stage  as shown in Fig.~\ref{fig:Profiler}. 
The stage moves the beam profiler repeatably between two points along the beam path separated by $\approx 82$~cm. 
The center of each beam is found using Gaussian fits to the profiler output at these two points. 
In order to ensure that the beams have Gaussian profiles and their centres can be determined accurately, the beams have been spatially filtered using an iris during the angle measurement.
Since the irises are fully open during the diffusion measurement, a potential systematic effect due to wavefront curvature arises which is quantified in Sec.~\ref{sec:results}.
This data is combined with measurements of the beam path to determine the angle between the excitation beams. 

\section{Experimental results} \label{sec:results} 

\subsection{Dependence of the decay time on angle and pressure}

The amplitude of the combined Rb population grating in each $m$ level is measured by applying a read-out pulse with a rise-time of $\approx67$~ns and a duration of $\approx50$~ns. 
The read-out pulse has an intensity of 120 mW/cm$^2$ optimized to maximize the scattered light signal.  
The read-out delay time relative to the excitation pulse is varied in a random sequence over $N\approx 1500$ delays. 
For each delay, the scattered light signal $E_{\rm C}(t)$ of Eq.~\ref{eq:diffsig}, is acquired by averaging four consecutive repetitions at the $\approx 1$~kHz repetition rate of the experiment.
We then acquire the background signal $E_{\rm BG}$ by turning off the $\vec k_2$ excitation pulse for the same delay time. 
The switching process requires $\approx 10$~ms. 
The signal $E_{\rm BG}$ is also acquired by averaging four consecutive repetitions.
We limit the number of consecutive traces to four because the phases of these electric fields are well correlated on timescales of $\approx 20$~ms.
This limitation is imposed by variations in the optical path length due to convection currents caused by the heated oven.
To improve statistics, each of the $N$ time delays is repeated four times in a random sequence of $4N$ acquisitions.
It takes $\approx150$~ms to transfer data and switch to a different delay time.
As a consequence, the acquisition time for the entire decay curve ranges from 5 min to 15 min.
This data acquisition procedure was modified only in nitrogen, for which the signal strength was the largest.
In this case the signal and background were acquired on the basis of 50 consecutive repetitions which were not repeated. 

For delay time $t$, signals $E_{\rm C}$  and $E_{\rm BG}$  are detected as a heterodyne beatnote at 80~MHz and mixed down to DC using the 80~MHz RF oscillator driving the AOMs.
The resulting in-phase, $V_{\rm p}$, and $\pi/2$-out-of-phase, $V_{\rm q}$, homodyne
signals are added in quadrature $\sqrt{V_{\rm p}^2 +V_{\rm q}^2}$ to obtain non-negative electric field amplitudes.
The associated phases $\phi_{\rm C}(t)$ and $\phi_{\rm BG}$ are  $\arctan(V_{\rm p}/V_{\rm q})$ when $V_{\rm q}>0$ and $\arctan(V_{\rm p}/V_{\rm q})+\pi$ when $V_{\rm q}<0$.
We have made it explicit that  $E_{\rm C}(t)$ and $\phi_{\rm C}(t)$ depend on delay time $t$
and, in practice, $0\le E_{\rm BG}\ll E_{\rm C}(t)$.
The non-negative background-subtracted (BS) signal amplitude is
\begin{equation}
\label{eq:backsub}
E_{\rm BS}(t)=  | E_{\rm C}(t) e^{i\phi_{\rm C}(t)}-E_{\rm BG} e^{i\phi_{\rm BG}}|
\end{equation}
as a function of delay time and $|z|$ is the absolute value of the complex argument $z$.

In the presence of amplitude and phase noise, the background signal can be described by probability density $p_{\rm BG}(E_{\rm BG},\phi_{\rm BG})= 
e^{-E_{\rm BG}^2/(2\sigma_{\rm BG}^2)}/(2\pi \sigma^2_{\rm BG})$.
This corresponds to a Gaussian distribution with a zero mean and a standard deviation $\sigma_{\rm BG}$ for $E_{\rm BG}e^{i\phi_{\rm BG}}$.
The acquired, background subtracted signal $E_{\rm aq}(t)$ becomes 
\begin{equation}
    E_{\rm aq}(t) =
      \int_{0}^\infty E_{\rm BG}{\rm d}E_{\rm BG} \int_0^{2\pi} {\rm d}\phi_{\rm BG}\,
     p_{\rm BG}(E_{\rm BG},\phi_{\rm BG})  E_{\rm BS}(t) \,.
\end{equation}
No closed form is available for this two-dimensional integral, but some algebra
shows that the background-subtracted  signal amplitude satisfies
\begin{equation}
    E_{\rm aq}(t) =
         \left\{\begin{array}{cl}
    \displaystyle \sqrt{\frac{\pi}{2}} \sigma_{\rm BG}\left[1+\frac{1}{4}\left(\frac{E_{\rm C}(t)}{\sigma_{\rm BG}}\right)^2\right]
       & {\rm for}\ E_{\rm C}(t)\ll \sigma_{\rm BG}\\
   \displaystyle    E_{\rm C}(t)\left[  1+\frac{1}{2} \left(\frac{\sigma_{\rm BG}}{E_{\rm C}(t)}\right)^2 \right]
       & {\rm for}\ E_{\rm C}(t)\gg \sigma_{\rm BG}
  \end{array}
  \right. \label{eq:aqui}
\end{equation}
omitting terms of higher order in $E_{\rm C}(t)/\sigma_{\rm BG}$ and $\sigma_{\rm BG}/E_{\rm C}(t)$ for the two limits, respectively.

Since we have not taken data for $E_{\rm C}(t)\ll \sigma_{\rm BG}$, we fit the measured $E_{\rm aq}(t)$
to
\begin{equation}
    E_{\rm aq}(t) = \sqrt{E_{\rm C}(t)^2 + \sigma_{\rm BG}^2}
    \label{eq:simplenoise}
\end{equation}
ensuring that we reproduce to first two terms of Eq.~\ref{eq:aqui} when $E_{\rm C}(t)\gg\sigma_{\rm BG}$, while still predicting a finite, although incorrect $E_{\rm aq}(t)$ for small $E_{\rm C}(t)$.
The standard deviation $\sigma_{\rm BG}$ is an adjusted parameter in addition to those
of $E_{\rm C}(t)$ in Eq.~\ref{eq:diffsig} with decay rate $1/\tau$ of Eqns.~\ref{eq:generaldecay} or \ref{eq:fulldecay}.

\begin{figure}
	\includegraphics[width=\linewidth]{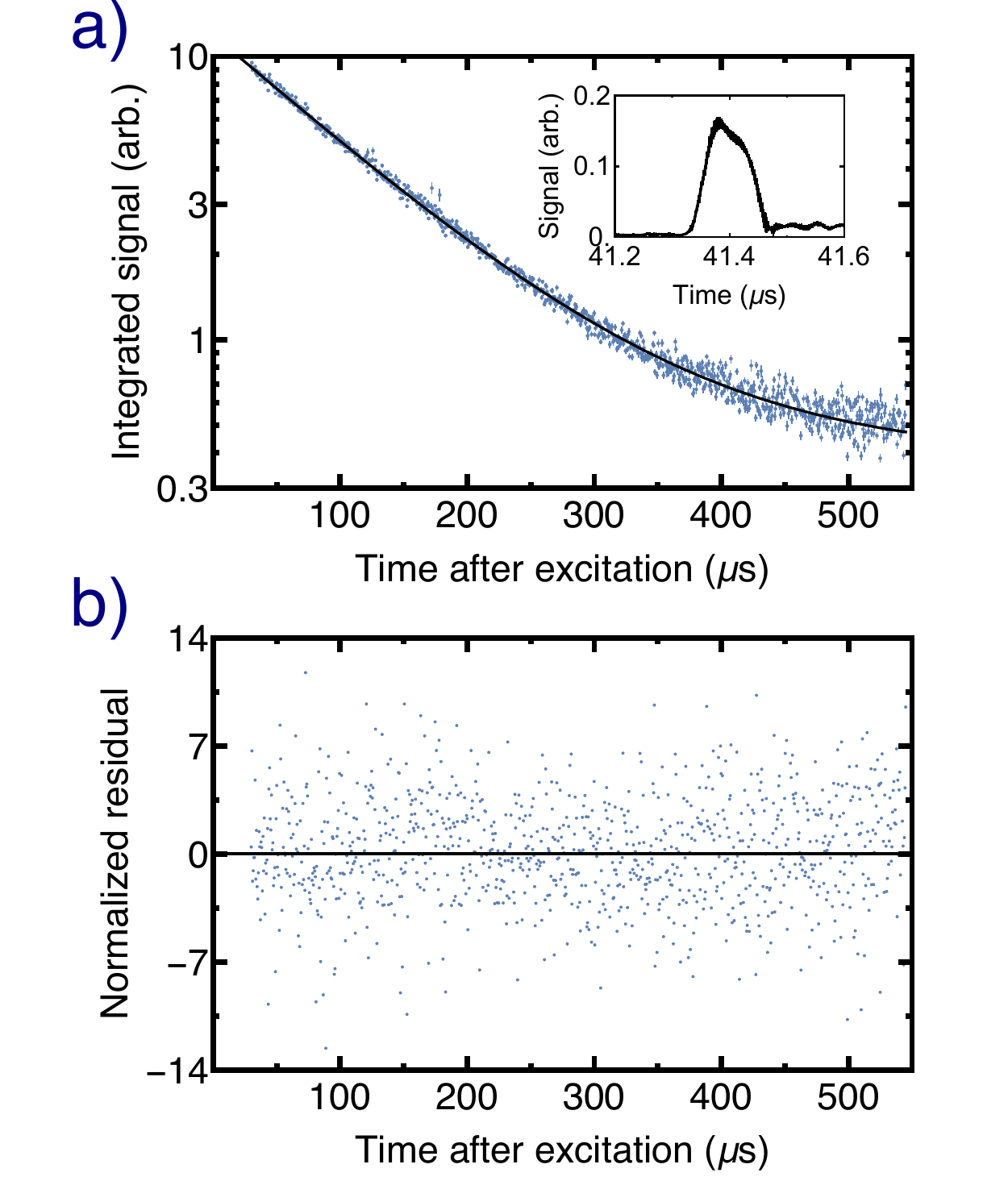}
	\caption{ (Color online)
a) Decay of the square of the amplitude of the optically pumped grating, the integrated signal ${\cal S}(t_i)$, with $341.70(5)$ Torr of neon buffer gas at $T=24.0(5)$~$^\circ$C as function of  time after excitation or delay time $t_i$. The excitation beams are aligned with an angle $\theta = 1.61(2)$ mrad.
Here, the parentheses represent one standard deviation of statistical and systematic uncertainty due to wavefront curvature as explained in the text.
A fit to $\sqrt{E_0^2\exp(-2t/\tau)+\sigma_{\rm BG}^2}$ gives a decay time constant of $123.0(4)$ $\mu$s.
Inset of panel a) shows the background-subtracted signal amplitude produced by a 50 ns readout pulse with a delay of 41.33 $\mu$s after the excitation pulses. The rise time of read out pulse is 67 ns. 
Such signals are integrated and squared to obtain each point in panel a).
b) Normalized residuals $({\cal S}(t_i)-E_{\rm aq}(t_i;E_0,\sigma_{\rm BG},\tau))/u(S_i)$ of our fit as a function of $t_i$.
}
	\label{fig:cleanExp}
\end{figure}

The inset of Figure~\ref{fig:cleanExp}a shows data for a typical example of the background-subtracted  signal amplitude, obtained for the delay time $t=41.33$~$\mu$s at a pressure of 342.70(5) Torr of neon buffer gas and $T=24.0(5)$ $^\circ$C. 
Each point on this curve has an error bar that represents the standard deviation of the 16 measurements.
The integrated area under such curves $S_i= S(t_i)$ is computed for each delay time $t_i$, $i=1,\dots, N$. The results are shown in Fig.~\ref{fig:cleanExp}a.
Each point $S_i$ has an error bar representing the standard uncertainty $u(S_i)=u(S(t_i))$ of the  integrated area, calculated as the quadrature sum of the error bars in the inset.

Using Eqns.~\ref{eq:diffsig} and \ref{eq:generaldecay}, we extract the decay time constant $\tau$  by minimizing chi square 
\begin{equation}
\label{eq:chisq}
    \chi^2=\sum_{i=1}^{N} \frac{(S_i-E_{\rm aq}(t_i;E_0,\sigma_{\rm BG},\tau))^2}{u^2(S_i)}\,,
\end{equation}
where for the acquired, background subtracted signal  $E_{\rm aq}(t;E_0,\sigma_{\rm BG},\tau)=\sqrt{E_{\rm C}(t)^2+\sigma^2_{\rm BG}}=\sqrt{E_0^2\exp(-2t/\tau)+\sigma_{\rm BG}^2}$ we have made the dependence on the  adjustable parameters $E_0$, $\sigma_{\rm BG}$, and $\tau$ apparent.

For the data shown in Fig.~\ref{fig:cleanExp}, where the non-exponential behavior for large delay times is clearly visible, we find that the standard uncertainty at a delay time, $u(S_i)$, generally underestimates the deviation from the fit value, {\it i.e.} the residual $S_i-E_{\rm aq}(t_i;E_0,\sigma_{\rm BG},\tau)$. 
This effect is quantified by the Birge ratio of the fit given by $\sqrt{\chi^2/(N-3)}=3.7$, which is significantly larger than 1 for our $N\gg 1$.
We attribute this inconsistency to the time-correlation between the consecutively acquired samples for each delay time, an effect that arises due to the phase drifts associated with the convection currents from the oven. In fact, from analyses of the fit residuals, we find that the residuals are proportional to $\sqrt{S_i}$, with a few exceptions in the nitrogen data at large angle for which the residuals are proportional to $S_i$.

Figure~\ref{fig:cleanExp}b
shows the {\it normalized} residuals $r_i=[S_i-E_{\rm aq}(t_i;E_0,\sigma_{\rm BG},\tau)]/u(S_i)$ as a function of delay time $t_i$. 
Their standard deviation $\sqrt{\langle r_i^2\rangle}=\sqrt{\chi^2/N}$  is clearly larger than 1.
More importantly, the figure shows that there exist no obvious correlations among the data.
In fact, the residuals are well represented by a normal distribution.
To account for the observed large residuals and to obtain a satisfactory fit, the standard uncertainty of decay time constant $\tau$ from the fit is multiplied by the Birge ratio.
We find that the typical Birge ratio for each gas is near 3.7, but it is as high as 5.6 in helium.

\begin{figure*}
	\centering
	\includegraphics[width=0.9\linewidth]{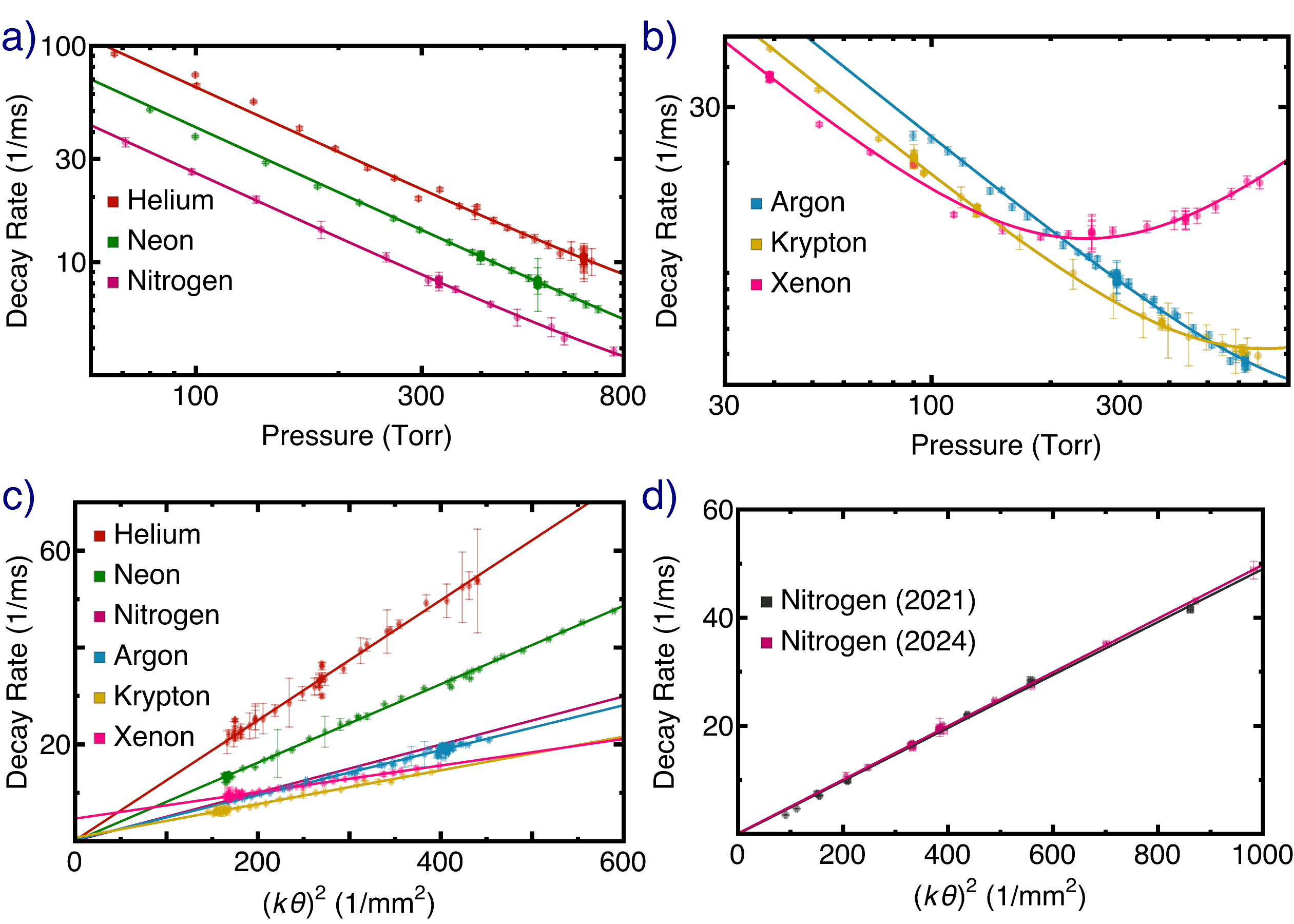}
\caption{
(Color online)
Data for naturally abundant Rb-$X$ systems at $T=24.0(5)$ $^\circ$C  with fits based on Eq.~\ref{eq:fulldecay} as described in the text.
a) Decay rates for helium, neon, and nitrogen buffer gasses as functions of buffer gas pressure
for angle $\theta=1.88$ mrad,  1.60 mrad, and 2.34 mrad, respectively.
b) Decay rates in argon, krypton, and xenon as functions of buffer gas pressure for $\theta=2.49$ mrad, 1.57 mrad, and 1.63 mrad, respectively. For all the angles reported here, the statistical uncertainty is small compared to the systematic uncertainty of $1.5\,\%$ which arises due to wavefront curvature as explained in text.
For ease of comparison, decay rates in panels a) and b) have been scaled to correspond to data taken
at a common angle of 2 mrad.
c) Decay rates as functions of $(k\theta)^2$ for helium, neon,  argon, krypton, and xenon
taken at buffer gas pressure 659.0(5) Torr, 526.73(1) Torr, 294.292(1) Torr, 90.27(1) Torr, and 38.953(1) Torr,
respectively. 
Only the best fit to the nitrogen data is shown in this panel. The nitrogen experimental data is displayed separately in panel d).
d) Decay rates as function of $(k\theta)^2$ for a nitrogen buffer gas at 325.97(2) Torr for this work 
and 552.4 Torr for our previous determination \cite{Diffusion}. 
Four data points from the previous determination fall outside the plot range.
For ease of comparison, decay rates in panels c) and d) have been scaled to correspond to data taken
at a common buffer gas pressure of 200 Torr.
The error bars in all panels represent one-standard-deviation  uncertainties obtained from
fits such as that shown in Fig.~\ref{fig:cleanExp}a.
}
	\label{fig:diffusionData}
\end{figure*}

Figure~\ref{fig:diffusionData} presents {\it all} measured decay rates $\tau$ and their standard uncertainties for the six inert buffer gasses and a large set of pressures $p$  and angles $\theta$.
These  measurements have been obtained at a temperature of $24.0(5)\,\degree$C and with the laser locked to the $F=3\rightarrow F'=3,4$ crossover peak in $^{85}$Rb.
Figures~\ref{fig:diffusionData}a and \ref{fig:diffusionData}b show observed decay rates as the pressure of each gas is varied at constant angle. 
Figures~\ref{fig:diffusionData}c and \ref{fig:diffusionData}d show the data as a function of the angle between laser beams at constant pressure. 
For a buffer gas we measure decay rates at $l$ pairs $(p,\theta)$. These data are fit to Eq.~\ref{eq:fulldecay} with positive fit parameters $\vsig_{\rm exp}$ and ${\cal Q}_{\rm exp}(T)$, and signed parameters $({\cal W}\Gamma_{\rm opt}p_{\rm ref})_i$, where index $i=1,\dots,m$ corresponds to data taken on different days
and models changes in residual light due to small deviations in the alignments of the AOMs.
For each buffer gas, the number of fit parameters $m+2$ is significantly smaller than the number of measured decay rates $l$. 

 To better visualize the $p$ and $\theta$ dependencies of the decay rates, we have plotted a modified decay rate in Fig.~\ref{fig:diffusionData}.
First, we  subtracted  $({\cal W}\Gamma_{\rm opt}p_{\rm ref})_i/p$ from each
measured $1/\tau$ using the corresponding fitted value $({\cal W}\Gamma_{\rm opt}p_{\rm ref})_i$.  
Then, for Figs.~\ref{fig:diffusionData}a and b, the resulting decay rates have been scaled to a common angle of $\theta'=2$ mrad. 
Here, the scaled rate $\left(1/\tau\right)'$ is determined by
\begin{equation}
\left(\frac{1}{\tau}\right)'=\left(\frac{1}{\tau}\right)\frac{g(p,\theta',T)}{g(p,\theta,T)},
\end{equation}
where $g(p,\theta,T)$ corresponds the first two terms on the right hand side of Eq.~\ref{eq:fulldecay}
using the fitted values for $\vsig$ and ${\cal Q}(T)$.
In a similar manner, the decay rates in Figures~\ref{fig:diffusionData}c and d have been scaled to a common pressure $p' = 200$ Torr.
As a result, a positive modified decay rate for $\theta\to0$ in Figures~\ref{fig:diffusionData}c and d indicates the presence of spin-exchange and spin-destruction collisions.

We  note that in Figs.~\ref{fig:diffusionData}a and \ref{fig:diffusionData}b the lowest pressures at which we can observe a signal are buffer-gas dependent.  They are determined by the limited efficiency of optical pumping for small collisional shifts and broadening as discussed in Sec.~\ref{sec:theory}.

Based on Eq.~\ref{eq:fulldecay}, we infer that the contributions from the Rb spin-exchange and spin-destruction collisions are small compared to that of Rb diffusion and residual light.
For the heavier argon, krypton and xenon gasses, the decay rates in Fig.~\ref{fig:diffusionData}b exhibit a component proportional to $p$, which is apparent at our larger pressures.
We interpret this trend as evidence of the influence of spin-exchange and spin-destruction collisions.
In Figs.~\ref{fig:diffusionData}c and \ref{fig:diffusionData}d
the linear increase with $(k\theta)^2$ is  solely due to diffusion.
While this dependence can be used to determine $D(T,p)$ even without the data in Figs.~\ref{fig:diffusionData}a and \ref{fig:diffusionData}b, the quality of the fits improves by including all data.

Figure~\ref{fig:diffusionData}d shows our nitrogen data as well as data from our previous determination in Ref.~\cite{Diffusion} as functions of angle $\theta$. The data from Ref.~\cite{Diffusion}, taken at a temperature of 50 $^\circ$C, has been scaled to 24.0(5) $^\circ$C using the procedure described in the caption of Table \ref{tab:compare}.
The fit to the nitrogen data gives $D(T,p)=0.131(3)$ cm$^2$/s at standard pressure, which agrees with the reported  $0.129(1)$ cm$^2$/s in Ref.~\cite{Diffusion}. 
We summarize the data from Fig.~\ref{fig:diffusionData} in Table~\ref{tab:diffusion} where the second column represents the values of $D(T,p)$ extracted from the fits to Eq.~\ref{eq:fulldecay} with statistical uncertainty shown in parentheses.
The second column of Table~\ref{tab:spinrate} shows the values of $\vsig$.

\subsection{Transit-time correction}

A transit time correction is estimated by simulating the signal decay based on two analytical models that 
attempt to account for the spatial expansion or diffusion of the Rb populations gratings, as described by 
Eq.~\ref{eq:diffusion}, beyond the extent of the read-out beam~\cite{myThesis}.
In both models the analytical signal $E(t)$ is given by the $\vec k_1-\vec k_2$ Fourier component of the product of population grating $\rho_{Fm_F}(\vec x,t)$ and the profile of the electric field amplitude of the read-out beam $E_{\rm RO}(\vec x)$. 

In the first model, we assume that the spatial amplitudes $B_{Fm_F}$ of the initial population gratings in Eq.~\ref{eq:population}
and $E_{\rm RO}(\vec x)$ are finite and positive on the same rectangular cuboid
and are zero elsewhere.
This cuboid reflects the spatial volume of the Gaussian laser beams, where 
the atomic transitions are saturated and a grating is formed, and the volume from which coherent light scattering can be detected.
The cuboid is assumed to be infinitely long along $(\vec k_1 + \vec k_2)/2$, our $y$ direction in Fig.~\ref{fig:grating}a), but has finite sizes $w_x= 3$ mm and $w_z = 4$ mm along directions $\vec k_1-\vec k_2$ and $\vec k_1 \times \vec k_2$, respectively, based on our simulations of optical pumping that show saturation.

This first model predicts that the  decay of the scattered field will be the real component of
\begin{widetext}
\begin{align}
	\begin{split}
	E_{\rm rect}(t)=&E_0\Bigg[
   \erf(d_z(t)/2)- \frac{2}{\sqrt{\pi}d_z(t)}[1-e^{-d_z^2(t)/4}]
         \Bigg]
         \Bigg[ -\frac{2}{\sqrt{\pi}d_x(t)}
    \left(1-e^{-d_x^2(t)/4}\cos(k\theta w_x) \right)\\
		    &\quad + e^{-\Delta^2(t)}\Bigg(\Bigg.1 - 2i \frac{\Delta(t)}{d_x(t)}\Bigg)\Bigg(\erf( d_x(t)/2-i\Delta(t))+\erf(i\Delta(t))\Bigg)\Bigg]\\
	\end{split}
    \label{eq:biggrect}
\end{align}
\end{widetext}
with the dimensionless time-dependent functions
\begin{eqnarray}
    \Delta(t) &=& k\theta \sqrt{D(T,p)t}\,,\\
    d_i(t) &=& w_i/\sqrt{D(T,p)t}
\end{eqnarray}
with $i=x$ and $z$.
Here, $\erf(z)$ is the error function of a complex variable $z$, $\erfi(z) =-i \erf(i z)$, $\real[z]$ and $\im[z]$ are the real and imaginary components of complex variable $z$.
Note that with $\erf(z)\to 1- \exp(-z^2)/\sqrt{\pi} z$ for $z\to\infty$ 
Eq.~(\ref{eq:biggrect}) approaches Eq.~(\ref{eq:diffsig}) when $w_i\to\infty$.

In the second model, we assume that the spatial amplitudes $B_{Fm_F}$ of the initial population gratings in Eq.~\ref{eq:population} and $E_{\rm RO}(\vec x)$ have the same Gaussian profile along orthogonal directions ${\vec k_1} - {\vec k_2}$ and ${\vec k_1} \times {\vec k_2}$.
The profiles of $B_{Fm_F}$ are assumed to be dictated by the laser intensity profile, which has a measured $1/e^2$ full width of $W_x=2.4$ mm and $W_z=2.4$ mm along these two directions, respectively. 
The profile of $E_{\rm RO}(\vec x)$ is the square root of the measured intensity profile and thus has a $1/e$ full width of $W_x=2.4$ mm and $W_z=2.4$ mm.
This model predicts a modified decay given by signal
\begin{widetext}
\begin{equation}
    E_{\rm Gauss}(t)=E_0
    \frac{W_x\sqrt{w^{**}_x}(t)}{w^*_x(t)}  \frac{W_z\sqrt{w^{**}_z}(t)}{w^*_z(t)}
    \exp\left(
     \frac{W_x^4-[w_x^*(t)]^2}{12 w_x^*(t)w_x^{**}(t)}W_x^2(k\theta)^2
    +\frac{4W_x^2-3w_x^{**}(t)}{3w_x^{**}(t)}(k\theta)^2D(T,p) t
     \right)\,,
\end{equation}
\end{widetext}
where ${w_i^{*}(t) = W_i^2 + 8D(T,p)t}$ and ${w_i^{**}(t) = W_i^2 + 8D(T,p)t}/3$ for $i=x$ and z.

The signals $E_{\rm rect}(t)$ and $E_{\rm Gauss}(t)$ are non-exponential in delay time $t$.
We then construct the acquired, background subtracted signal as
\begin{equation}
    \sqrt{ [E_\alpha(t)e^{-\vsig pt/k_{\rm B}T}]^2 + {\rm BG}_{\rm sim}^2 }\,,
    \label{eq:transitnoise}
\end{equation}
where $\alpha={\rm rect}$ or Gauss, in analogy to the expression for the signal in Eq.~(\ref{eq:simplenoise}).
Here, factor $\exp(-\vsig pt/k_{\rm B}T)$ accounts for the effects of spin-destruction collisions and $\sigma_{\rm BG\mhyphen sim}$ represents the phase and amplitude noise of the heterodyne detection.  
We do not directly fit our experimental data to these model signals, but instead generate discrete noisy datasets based on Eq.~(\ref{eq:transitnoise}) for each pair of $(p,\theta)$ for which a time trace was acquired. 
To generate these datasets, we use estimates for ${\cal Q}$ and $\vsig$, and assume a one-dimensional normal distribution for ${\rm BG}_{\rm sim}$ with a mean of zero and width given by $\sigma_{\rm BG}$ for that time trace.
We take the average of 16 of these simulated time traces and fit the averaged datasets to the  Eq.~(\ref{eq:simplenoise}) with an exponential signal $E_{\rm C}(t)$ and obtain a modified rate $\tau'$ and $\sigma'_{\rm BG}$ for each time trace.
Next, we fit the values of $\tau'$ as a function of $\theta$ and $p$ to Eq.~(\ref{eq:fulldecay}) to obtain output values for ${\cal Q}'$ and for $\vsig'$, which are now systematically offset from ${\cal Q}$ and $\vsig$ due to the effects of transit time and the heterodyne noise. 

We adjust our input values for ${\cal Q}$ and $\vsig$ in Eq.~(\ref{eq:transitnoise}) until we arrive at a range of simulated values of ${\cal Q}'$ and $\vsig'$ that match our experimental observations of ${\cal Q_{\rm exp}}$ and $\vsig_{\rm exp}$ within the experimental error bar.
The mean of the estimates which satisfy this condition are taken as the corrected values of ${\cal Q}$ and $\vsig$.
The uncertainty in the corrected values has been evaluated by simulating a library of decay curves obtained by varying  input parameters $p$, $\theta$, $w_i$, $W_i$, ${\cal Q}$, and $\vsig$ based on their uncertainties. 

The transit time corrected values of $D(T,p)$ for each model are displayed in columns 3 and 4 of Table~\ref{tab:diffusion}. 
The uncertainty in these values has been added in quadrature to the statistical error in column one of Table~\ref{tab:diffusion} to obtain the error values reflected in columns three and four. 
Similarly, the transit time corrected values of $\vsig$ and their associated uncertainties are displayed in columns 3 and 4 of Table~\ref{tab:spinrate}.

We observe that the corrected values of $D(T,p)$ in columns 3 and 4 of Table~\ref{tab:diffusion} are larger than the uncorrected values. 
This behavior arises from the functional form of the non-exponential decay curves predicted by the models, which exhibit a more rapid decay for short delay times, and a less rapid decay for longer decay times.

Column 5 of Table~\ref{tab:diffusion} represents the final experimental value for $D(T,p)$ given by the weighted average of the results of the two models for the transit time correction. 
We observe that the Gaussian model predicts a very small correction, whereas the correction is substantial in the rectangular model. 
Since we do not know which of these models is correct, we choose the conservative option of reporting the weighted average.
The error in this value is computed as the quadrature sum of the error in the rectangular model, the error in the Gaussian model, and half the difference between the corrected value given by these two models.
Column 5 of Table~\ref{tab:spinrate} shows the final experimental value for $\vsig$ using the same procedure.

A second systematic effect relates to the curvature of the laser wavefronts that are incident on the aperture used to spatially filter the excitation beams in order to measure angle $\theta$.
This effect does not change the fitted value of $\tau$ but does introduce an additional fractional
uncertainty  of 1.5\,\%, which is added in quadrature to the statistical uncertainty in column two of Table~\ref{tab:diffusion}, and reflected in columns three through five.
To obtain this uncertainty, we translate the circular aperture across the excitation beam profiles and find that the measured angle varies by as much as 130 $\mu$rad per mm of translation. 
As the maximum uncertainty in the placement of the aperture is 0.2 mm, the maximum systematic uncertainty in the angle is 26 $\mu$rad. 
The impact of this effect on $D(T,p)$ as a function of $\theta$ and $p$ is as large as 1.5\,\%.

Finally, we compare our the weighted experimental values reported in column five of Table~\ref{tab:diffusion}, to those predicted by the quantum theory.
We find deviations from the experimental determinations of 0\,\% to 12\,\% for the six gases, all of which are in agreement within experimental errors.

\begin{table*}
	\caption{Measured Rb-inert gas diffusion coefficients $D(T,p)$ for naturally abundant Rb and our calculated $^{87}$Rb-inert gas diffusion coefficients at standard atmospheric pressure and $T=24.0(5)^\circ$C. Values and uncertainties in parenthesis are reported in cm$^2$/s.
 The diffusion coefficients with their statistical uncertainties inferred from fits are reported  in column two.
Columns three and four show $D(T,p)$ corrected for transit time effects using two models as explained in the text. 
Column five shows our recommended value, a weighted average of columns three and four, with errors in parentheses calculated as described in the text.
Diffusion coefficients for $^{87}$Rb-$X$ systems calculated using the quantum mechanical model are shown in the last column.
 }
	\label{tab:diffusion}
\renewcommand{\arraystretch}{1.1}
\begin{tabular}{|c|@{\ \ }l@{\ }|@{\ \ }l@{\ \ }l@{\ \ }l@{\ \ }|l@{\ \ }|}
	\hline
	Gas & \multicolumn{1}{c|@{\ \ }}{$D(T,p)$} &   \multicolumn{1}{c}{$D(T,p)$}  &   \multicolumn{1}{c}{$D(T,p)$}  &   \multicolumn{1}{c|}{$D(T,p)$} & \multicolumn{1}{c|}{$D(T,p)$}  \\
        &          &  \multicolumn{1}{c}{corrected}  &  \multicolumn{1}{c}{corrected}  &  \multicolumn{1}{c|}{recommended} &  \multicolumn{1}{c|}{quantum}\\
        &           &   \multicolumn{1}{c}{(rect.)} & \multicolumn{1}{c}{(Gaus.)} & & \\
        \hline
	He                     & 0.33(2)   & 0.34(4)   & 0.33(2)   & 0.33(5)      & 0.3768(16) \\
	Ne                     & 0.213(7)  & 0.217(11) & 0.213(8)  &  0.214(14)     & 0.212(3)   \\
	N$_2$                  & 0.131(3)  & 0.134(5)  & 0.131(4)  & 0.132(7)      & 0.1257(3)  \\
    N$_2$~\cite{Diffusion} & 0.129(1)  & 0.130(3)  & 0.129(2)  & 0.129(4)      &            \\
	Ar                     & 0.122(5)  & 0.124(7)  & 0.122(5)  & 0.123(9)      & 0.1329(13) \\
	Kr                     & 0.092(5)  & 0.096(7)  & 0.092(5)  &  0.093(9)     & 0.0960(11) \\
	Xe                     & 0.072(3)  & 0.074(3)  & 0.072(3)  &  0.073(4)     & 0.0709(5)  \\
	\hline
\end{tabular}
\end{table*}

From the compilation of $\vsig$ in Table~\ref{tab:spinrate}, we find our measurements show increasing positive values as a function of the mass of the buffer gas.
However, the values are also consistent with zero. This is likely due to the fact that we operate under conditions in which diffusion dominates other decay rates and thus limits our sensitivity to $\vsig$.
For Kr and Xe our values deviate significantly from
previous measurements at a similar temperature of $27\,\degree$C by Ref.~\cite{Bouchiat1972}. 
While we do not have an adequate explanation, we note that the values of the diffusion coefficients are not impacted by the values of $\vsig$ and vice-versa. 
As shown in Eq.~\ref{eq:fulldecay}, this is because the decay rate due to diffusion scales as $p^{-1}$ and $\theta^2$, while the decay rate due to $\vsig$ scales only as $p$, with no known coupling between these two effects.
We speculate that the discrepancy between our values of $\vsig$ and previous measurements arises from the fact that the measurements in Ref.~\cite{Bouchiat1972} were carried out using highly polarized Rb in the presence of a small, 1 Torr concentration of buffer gas leading
to longer timescales $\tau$ and thus greater sensitivity to $\vsig$. 
Measurements of $\vsig$ at a higher temperature~\cite{Walker2001} are also discrepant with our values when scaled to the same temperature.

\begin{table*}
\caption{
Measured $\vsig/k_{\rm B}T$ in unit s$^{-1}$Torr$^{-1}$ at $T=24.0(5)$ $^\circ$C  for spin-exchange and spin-destruction processes of trace amounts of naturally abundant Rb in inert buffer gasses $X$. 
The fitted values are shown in column two with their statistical uncertainties in parenthesis.
Columns three and four show values after correcting for systematic errors due to transit time using two models as described in the text.
Column five shows our recommended value, a weighted average of columns three and four as explained in text.
Our rate coefficients are compared to those measured by Ref.~\cite{Bouchiat1972} in column six where available. 
Note that $\vsig/(k_{\rm B}T)$ expressed in unit s$^{-1}$Torr$^{-1}$ is equivalent to $\vsig$ expressed in unit $3.08 \times 10^{-17}$ cm$^3$/s at $24\,\degree$C.
 }
	\label{tab:spinrate}
\renewcommand{\arraystretch}{1}
	\begin{tabular}{|c|c|ccc|c|}
		\hline
		Gas   & $\vsig/(k_{\rm B}T)$  &  $\vsig/(k_{\rm B}T)$  & $\vsig/(k_{\rm B}T)$ & $\vsig/(k_{\rm B}T)$ &$\vsig/(k_{\rm B}T)$\\
       $X$    &  &  corrected & corrected & recommended &\cite{Bouchiat1972}\\
              &  & (rect.) & (Gaus.) &  &\\
		\hline
		He    & 0(11)  & 0(11)  & 0(11) &   0(16)   & -         \\
		Ne    & 0(2)   & 0(2)   & 0(2)  &   0(3)   & -         \\
		N$_2$ & 1(3)   & 0(4)   & 1(3)  &   0(5)   & -         \\
		Ar    & 2(2)   & 2(3)   & 2(2)  &   2(4)   & 0.95(3)   \\
		Kr    & 4(3)   & 4(4)   & 4(3)  &   4(5)   & 33.2(1.5) \\
		Xe    & 23(15) & 22(21) & 23(15)&    23(26)  & 185(10)   \\
		\hline
\end{tabular}
\end{table*}

\subsection{Discussion and applications}

\begin{figure}
	\includegraphics[width=\linewidth, trim=0 30 0 0,clip]{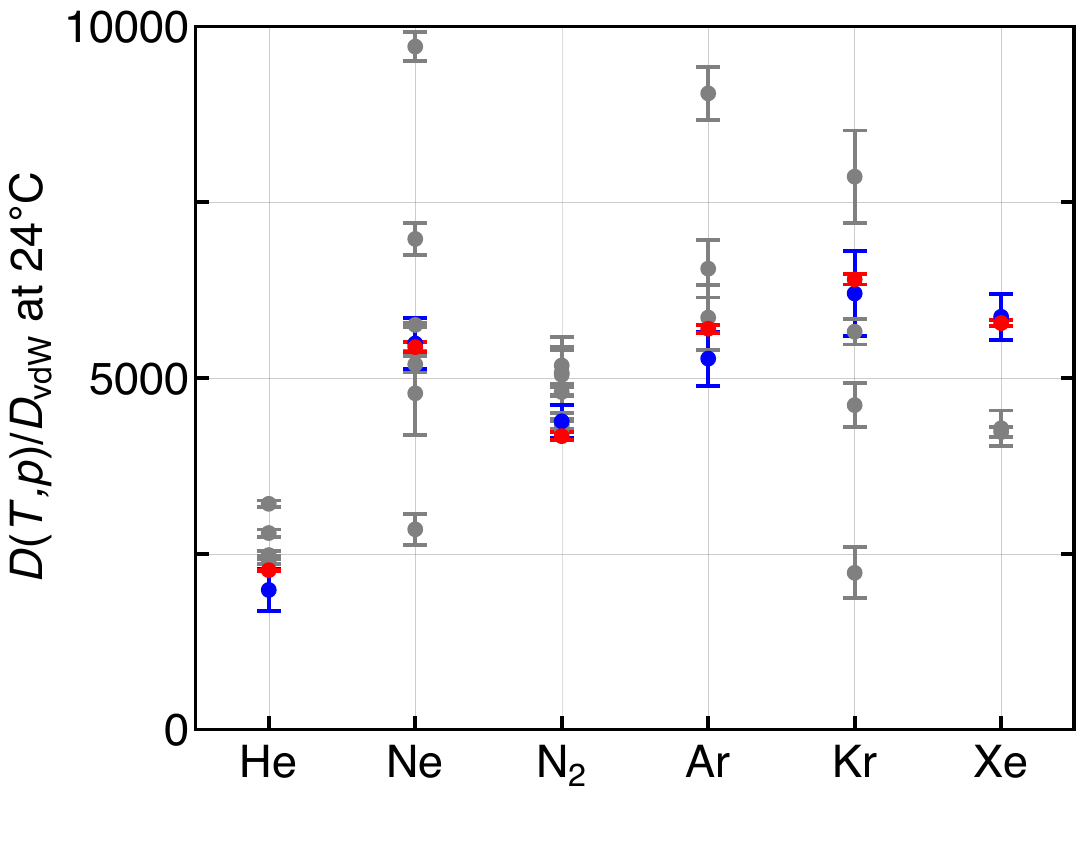}
	\caption{(Color online)
 Ratio $D(T,p)/D_{\rm vdW}$, where $D_{\rm vdW}=\beta_6v_6=\hbar/\mu$, for all six naturally abundant Rb+$X$ systems.
 Filled blue circles with error bars show this ratio based on the experimental determinations presented in this paper. 
 The ratios based on quantum theoretical values for $^{87}$Rb+$X$ systems are shown as red circles with error bars.
 We also display the same ratios with respect to other experimental results (in Table~\ref{tab:compare}) as gray circles with error bars, where  we assume  a standard uncertainty of 1 in the last significant digit for publications that did not supply an uncertainty. 
 The diffusion coefficients $D(T,p)$ are at standard atmospheric pressure and measured at or scaled to $T=24.0(5)^\circ$C.
 }
	\label{fig:vdWPlot}
\end{figure}

Our results allow us to make detailed comparisons between theory and experiment involving six distinct Rb-$X$ systems.
Figure~\ref{fig:vdWPlot} shows  $D(T,p)$ for all six buffer gas systems with data from our experiments, our values from quantum mechanical calculations,
and results of other experiments (see Table~\ref{tab:compare}).
Here, the vertical axis is dimensionless and represents the ratio $D(T,p)/D_{\rm vdW}$, where the  van-der-Waals diffusion coefficient $D_{\rm vdW}=\beta_6 v_6$, with relative van-der-Waals velocity $v_6 = \sqrt{2E_6/\mu}$ so that $D_{\rm vdW}=\hbar/\mu$, which only depends on the reduced mass $\mu$ of the system. 
This ratio, with most values between 2500 and 7500, allows us to compare the six buffer gas systems on the same vertical scale.
Our experimental values for $D(T,p)$ agree with the quantum mechanical calculations across all of the different gases. 
We also find a similar systematic variation in both experiment and theory as a function of the mass of the buffer gas.
The other historical measurements of $D(T,p)$, however, show a much larger spread of values relative to the theoretical predictions.

\begin{figure}
	\includegraphics[width=\linewidth]{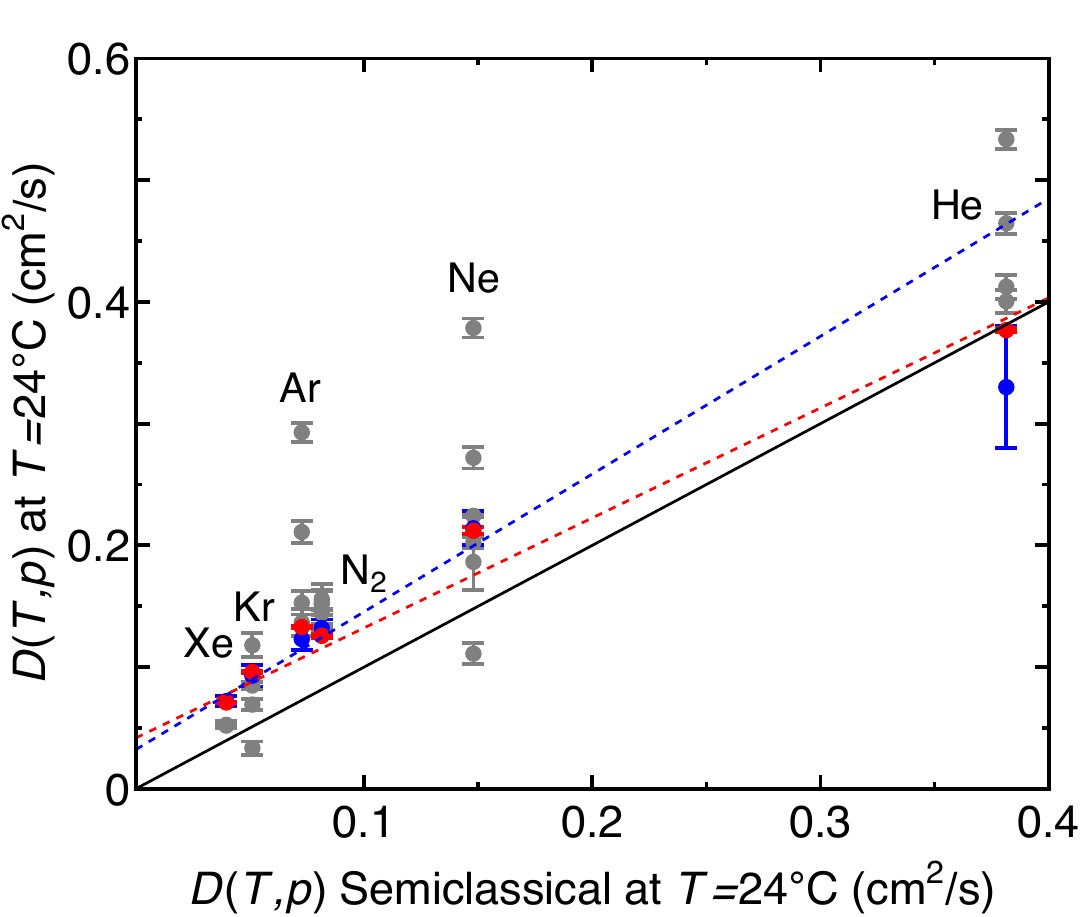}
	\caption{(Color online)
 Measured $D(T,p)$ for six natural abundance Rb+$X$ systems and theoretical $^{87}$Rb+$X$ diffusion coefficients from quantum mechanical simulations at standard atmospheric pressure and $T=24$ $^\circ$C plotted as functions of the corresponding semi-classical values for $D(T,p)$.
	Blue and red markers with error bars representing standard uncertainties correspond to our experimental and quantum theoretical values, respectively. 
    Gray markers correspond to previous measurements found in Table~\ref{tab:compare},
    where for publications that did not supply an uncertainty budget we assume a standard uncertainty of 1 in the last significant digit.
     The dotted blue and red  lines correspond to linear least-squares fits to the blue and red markers, respectively.
     The solid black line represents the semiclassical value for $D(T,p)$}
	\label{fig:scaling}
\end{figure}

In the semiclassical formalism, an analytical expression for $D(T,p)$ follows from the expression for the thermally averaged cross section ${\cal A}(T)$ in Eqs.~\ref{eq:PInvariant} and \ref{eq:AvdW}. 
It is then instructive, as shown in Fig.~\ref{fig:scaling}, to plot the measured $D(T,p)$ along with those from our quantum theoretical calculations as functions of $D(T,p)$ from the semiclassical treatment at one standard atmospheric pressure and $T= 24\,\degree$C. 
The graph shows that the experimental and semiclassical values agree to within a factor of two for all inert 
buffer gasses as already discussed for nitrogen in Fig.~\ref{fig:DiffRbN2}. It is worth noting that
the semiclassical diffusion coefficients depend on both $\mu$ and
the van-der-Waals coefficient $C_6$. In fact, we recall that the semiclassical $D(T,p)$ scales as $\mu^{-1/2}C_6^{-1/3}$.

Linear fits to our experimental and quantum diffusion coefficients with respect to the
semiclassical values are also shown in Fig.~\ref{fig:scaling}.
The slopes for the experimental and quantum predictions are unit-less and have values of $1.13(16)$ and $0.90(7)$ respectively, both of which differ from the semiclassical slope of unity.

In addition to the tabulated systematic uncertainties we have examined additional systematic effects that can affect the experimental diffusion coefficients.
Firstly, we have investigated the role of broadband background light due to, for example, amplified spontaneous emission (ASE) from our tapered amplifier by spatially filtering the excitation and readout beams using an optical fiber. 
Reference~\cite{SPIEDefense} has shown that the spectral intensity of  ASE can be reduced by a factor of 5.8 by using an  optical fiber. 
However, spatial filtering  did not affect the values for $D(T,p)$. 

We have also considered two other systematic effects that can potentially impact the diffusion coefficients.
These are i) spin decoherence due to the formation of Rb + $X$ van der Waals molecules~\cite{Bouchiat1972,Franz1976,Hartmann1970,HapperSpin,Saam92,Walker2001} and
ii) the redistribution of light due to radiation trapping, including the role of quenching and mixing collisions.
These effects can modify the formation of the population grating  by depopulating the excited state and redistributing light and result in an inverse pressure dependence.
The first of these two effects is known to be small at the typical, $>10$ Torr buffer gas pressures used in this work~\cite{HapperSpin}. 
Secondly, based on reference~\cite{Holstein}, we have determined that the lifetime of the excited state of Rb within the cylindrical excitation volume is only extended by a factor of$\approx3$ due to radiation trapping. This timescale is three orders of magnitude smaller than the timescale of diffusion and is therefore unlikely to play a role in the observed decays. 
As collisional quenching and mixing are only relevant  when radiation trapping is prominent, these effects can be ruled out for the same reason. 
Additionally, Ref.~\cite{Speller1979} suggests that collisional quenching rates in noble gases are of order 200~Hz at 760~Torr, which is too small to affect our measurements even at the upper limit of our experimental pressure range. 
We conclude that these processes have not influenced our measurements.

We conclude this subsection by describing a possible application of our setup, namely, the realization of a quantum pressure sensor that depends on the intrinsic properties of atomic interactions. 
Since experiment and theory are in agreement, it is possible to solve 
Eqs.~\ref{eq:definition} and \ref{eq:PInvariant} directly for $p$ by first determining ${\cal Q}(T)$ by recording decay curves over a range of angles and using the calculated value of ${\cal A}(T)$.
By following this procedure, one can infer the absolute value of the pressure of a buffer gas $X$ in a Rb-$X$ mixture, over the effective pressure range of the technique.
In this manner our technique can serve as a basis for a quantum pressure sensor. 
Such an effort would complement other techniques that have been recently developed to realize pressure sensors using ultracold atoms under high vacuum conditions~\cite{Turlapov2016,Tiesinga2017,Shen2020,Comparison2022}.

\section{\label{sec:conc}Concluding remarks}

We have presented unified measurements of diffusion coefficients near room temperature  along with theoretical predictions using quantum, semiclassical, and classical models for six inert buffer gases and trace amounts of natural abundance rubidium using a single experimental apparatus.
Our experiment relies on creating spatial population gratings in each $m$-level of the rubidium gas and measuring its
decay or diffusion using two nearly parallel optical laser beams. The decay rate is proportional to the diffusion coefficient and
the square of the small angle between the laser beams.

By accounting for systematic effects we have resolved discrepancies between the experimental and theoretical diffusion coefficients for six buffer gases, which provides the basis for a pressure sensor.
A practical realization of this idea  requires confirmation of the predicted temperature dependence of $D(T,p)$.
A new round of experiments will be able to realize greater precision as the optical pumping simulations in Fig~\ref{fig:mLevels} suggest that an increase in the signal to noise ratio may be possible by tuning the read-out and excitation pulses to be resonant with separate ground state hyperfine levels. 
Systematic effects due to signal offsets, as shown in Fig.~\ref{fig:cleanExp}, can be avoided by using a detection system consisting of a photomultiplier tube  with an electronic gate and an AOM shutter as in Ref.~\cite{GehrigPendellosung}.
It may also be possible to reduce the effect of wavefront curvature on the systematic error in the angle $\theta$ by spatially filtering the excitation and readout beams using optical fibers.
A better description of the effects of residual laser light on the rubidium atoms can be based explicitly on the line-shape function as described by the last term in Eq.~\ref{eq:generaldecay} instead of the approximation described by the last term in Eq.~\ref{eq:fulldecay}. 
Such an improved description, which must include all magnetic sublevels and their relative populations, may allow us to reduce the error associated with this effect.

\section{\label{sec:acks}Acknowledgements}

We acknowledge helpful discussions with Krishna Myneni, US Army DEVCOM Aviation and Missile Center, Redstone Arsenal, and Brian Saam, Washington State University. We thank James Whiteway, York University for the loan of an optical table. We also thank Brynle Barrett and Dennis Tokaryk, University of New Brunswick for the supply of xenon and Greg Koyanagi, York University for the supply of neon. 
This  work was  supported  by  the  Canada  Foundation  for Innovation,  the Ontario  Innovation  Trust,  the  Ontario  Centers  of Excellence, the Natural Sciences and Engineering Research Council of Canada, York University and the Helen Freedhoff Memorial fund. 

\bibliography{apssamp}

@PREAMBLE{
 "\providecommand{\noopsort}[1]{}" 
 # "\providecommand{\singleletter}[1]{#1}%" 
}

@article{Xantheas2025,
    author = {Perko, Joseph R. and Xantheas, Sotiris S.},
    title = {Collision integrals within the Chapman–Enskog theory for a generalized Lennard-Jones potential},
    journal = {The Journal of Chemical Physics},
    volume = {162},
    number = {3},
    pages = {034113},
    year = {2025},
    month = {01},
    issn = {0021-9606},
    doi = {10.1063/5.0244532},
    url = {https://doi.org/10.1063/5.0244532},
}

@article{Comparison2022,
    author = {Ehinger, Lucas H. and Acharya, Bishnu P. and Barker, Daniel S. and Fedchak, James A. and Scherschligt, Julia and Tiesinga, Eite and Eckel, Stephen},
    title = "{Comparison of two multiplexed portable cold-atom vacuum standards}",
    journal = {AVS Quantum Science},
    volume = {4},
    number = {3},
    pages = {034403},
    year = {2022},
    month = {07},
    issn = {2639-0213},
    doi = {10.1116/5.0095011},
    url = {https://doi.org/10.1116/5.0095011}}

@article{Bernheim62,
    author = {Bernheim, Robert A.},
    title = "{Spin Relaxation in Optical Pumping}",
    journal = {J. Chem. Phys.},
    volume = {36},
    number = {1},
    pages = {135-140},
    year = {1962},
    month = {01},
    issn = {0021-9606},
    doi = {10.1063/1.1732283},
    url = {https://doi.org/10.1063/1.1732283}
}

@article{Chrapkiewicz2014,
title = {How to measure diffusional decoherence in multimode rubidium vapor memories?},
journal = {Opt. Commun.},
volume = {317},
pages = {1-6},
year = {2014},
doi = {https://doi.org/10.1016/j.optcom.2013.12.020},
url = {https://www.sciencedirect.com/science/article/pii/S0030401813011577},
author = {Radosław Chrapkiewicz and Wojciech Wasilewski and Czesław Radzewicz}
}

@article{Aymar69,
	author = {{Aymar, M.} and {Bouchiat, M.A.} and {Brossel, J.}},
	title = {\'{E}tude exp\'erimentale de la relaxation du rubidium en pr\'esence d'h\'elium},
	DOI= "10.1051/jphys:01969003008-9061900",
	url= "https://doi.org/10.1051/jphys:01969003008-9061900",
	journal = {J. Phys. France},
	year = {1969},
	volume = {30},
	number = {8-9},
	pages = "619-629",
}

@article{HapperCoating,
  title = {Modification of glass cell walls by rubidium vapor},
  author = {Ma, J. and Kishinevski, A. and Jau, Y.-Y. and Reuter, C. and Happer, W.},
  journal = {Phys. Rev. A},
  volume = {79},
  issue = {4},
  pages = {042905},
  numpages = {5},
  year = {2009},
  month = {Apr},
  publisher = {American Physical Society},
  doi = {10.1103/PhysRevA.79.042905},
  url = {https://link.aps.org/doi/10.1103/PhysRevA.79.042905}
}

@article{GehrigPendellosung,
  title = {Role of optical channeling in contrast enhancement of echo interferometers},
  author = {Carlse, Gehrig and Randhawa, Jaskaran and Ramos, Eduardo and Vacheresse, Thomas and Pouliot, Alex and Carew, Adam C. and Kumarakrishnan, A.},
  journal = {Phys. Rev. A},
  volume = {109},
  issue = {4},
  pages = {043307},
  numpages = {8},
  year = {2024},
  month = {Apr},
  publisher = {American Physical Society},
  doi = {10.1103/PhysRevA.109.043307},
  url = {https://link.aps.org/doi/10.1103/PhysRevA.109.043307}
}

@article{AbdelHafiz2017,
   title={A high-performance {R}aman-{R}amsey {C}s vapor cell atomic clock},
   volume={121},
   url={http://dx.doi.org/10.1063/1.4977955},
   number={10},
   journal={J. Appl. Phys.},
   publisher={AIP Publishing},
   author={Abdel Hafiz, Moustafa and Coget, Grégoire and Yun, Peter and Guérandel, Stéphane and de Clercq, Emeric and Boudot, Rodolphe},
   year={2017}
}

@Article{Lvovsky2009,
author={Lvovsky, Alexander I.
and Sanders, Barry C.
and Tittel, Wolfgang},
title={Optical quantum memory},
journal={Nature Photonics},
year={2009},
month={Dec},
day={01},
volume={3},
number={12},
pages={706-714},
issn={1749-4893},
doi={10.1038/nphoton.2009.231},
url={https://doi.org/10.1038/nphoton.2009.231}
}

@article{Steck85,
author = {Steck, D.},
year = {2023},
month = {9},
pages = {},
url={https://steck.us/alkalidata/},
journal={https://steck.us/alkalidata/},
note = {{R}ubidium-85 $\mathrm{D}$ Line Data (Version 2.3.2) }
}

@book{bermanTextbook,
  title={Principles of laser spectroscopy and quantum optics},
  author={Berman, Paul R and Malinovsky, Vladimir S},
  year={2011},
  publisher={Princeton University Press}
}

@article{BermanRateEq,
  title = {Rate equations between electronic-state manifolds},
  author = {Berman, P. R. and Rogers, G. and Dubetsky, B.},
  journal = {Phys. Rev. A},
  volume = {48},
  issue = {2},
  pages = {1506--1513},
  numpages = {0},
  year = {1993},
  month = {Aug},
  publisher = {American Physical Society},
  doi = {10.1103/PhysRevA.48.1506},
  url = {https://link.aps.org/doi/10.1103/PhysRevA.48.1506}
}

@article{Steck87,
author = {Steck, D.},
year = {2023},
month = {9},
pages = {},
url={https://steck.us/alkalidata/},
journal={https://steck.us/alkalidata/},
note = {{R}ubidium-87 $\mathrm{D}$ Line Data (Version 2.3.2) }
}

@article{Diffusion-Erratum,
  title = {Erratum: Accurate determination of an alkali-vapor–inert-gas diffusion coefficient using coherent transient
emission from a density grating [$\textrm{P}$hys. $\textrm{R}$ev. $\textrm{A}$ 103, 023112 (2021)]}, 
  author = {Pouliot, Alexander and Carlse, Gehrig and Beica, Hermina C. and Vacheresse, Thomas and Kumarakrishnan, A. and Shim, Unyob and Cahn, S.B. and Turlapov, Andrey and Sleator, Tycho},
  journal = {Submitted to Phys. Rev. A},
  year = {2024},
  publisher = {American Physical Society},
}

@book{Corney,
  title={Atomic and Laser Spectroscopy},
  author={Corney, A.},
  isbn={9780198511380},
  lccn={77372597},
  series={Oxford science publications},
  url={https://books.google.ca/books?id=KgVowQEACAAJ},
  year={1977},
  publisher={Clarendon Press}
}

@article{Diffusion,
  title = {Accurate determination of an alkali-vapor--inert-gas diffusion coefficient using coherent transient emission from a density grating},
  author = {Pouliot, A. and Carlse, G. and Beica, H. C. and Vacheresse, T. and Kumarakrishnan, A. and Shim, U. and Cahn, S. B. and Turlapov, A. and Sleator, T.},
  journal = {Phys. Rev. A},
  volume = {103},
  issue = {2},
  pages = {023112},
  numpages = {10},
  year = {2021},
  month = {Feb},
  publisher = {American Physical Society},
  doi = {10.1103/PhysRevA.103.023112},
  url = {https://link.aps.org/doi/10.1103/PhysRevA.103.023112},
  note= {Corrected in ~\cite{Diffusion-Erratum}}
}

@article{Miller2016,
title = {High pressure line shapes of the {R}b {D}1 and {D}2 lines for $^4${H}e and $^3${H}e collisions},
journal = {J. Quant. Spectrosc. Radiat. Transfer},
volume = {184},
pages = {118-134},
year = {2016},
issn = {0022-4073},
doi = {https://doi.org/10.1016/j.jqsrt.2016.06.027},
url = {https://www.sciencedirect.com/science/article/pii/S0022407316301170},
author = {Wooddy S. Miller and Christopher A. Rice and Gordon D. Hager and Mathew D. Rotondaro and Hamid Berriche and Glen P. Perram}
}

@article{Rothe1959,
    author = {Rothe, Erhard W. and Bernstein, Richard B.},
    title = "{Total Collision Cross Sections for the Interaction of Atomic Beams of Alkali Metals with Gases}",
    journal = {J. Chem. Phys.},
    volume = {31},
    number = {6},
    pages = {1619-1627},
    year = {1959},
    month = {08},
    issn = {0021-9606},
    url = {https://doi.org/10.1063/1.1730662}
}

@article{RabiCrossSection,
  title = {Effective Collision Cross Sections of the Alkali Atoms in Various Gases},
  author = {Rosin, Seymour and Rabi, I. I.},
  journal = {Phys. Rev.},
  volume = {48},
  issue = {4},
  pages = {373-379},
  numpages = {0},
  year = {1935},
  month = {Aug},
  publisher = {American Physical Society},
  doi = {10.1103/PhysRev.48.373},
  url = {https://link.aps.org/doi/10.1103/PhysRev.48.373}
}

@article{BlankWeeksKedziora2012,
    author = {Blank, L and Weeks, David E. and Kedziora, Gary S.},
    title = "{M+Ng potential energy curves including spin-orbit coupling for M = K, Rb, Cs and Ng = He, Ne, Ar}",
    journal = {J. Chem. Phys.},
    volume = {136},
    number = {12},
    pages = {124315},
    year = {2012},
    month = {03},
    issn = {0021-9606},
    url = {https://doi.org/10.1063/1.3696377}
}

@article{Bouchiat1972,
    author = {Bouchiat, M. A. and Brossel, J. and Pottier, L. C.},
    title = "{Evidence for Rb‐Rare‐Gas Molecules from the Relaxation of Polarized Rb Atoms in a Rare Gas. Experimental Results}",
    journal = {J. Chem. Phys.},
    volume = {56},
    number = {7},
    pages = {3703-3714},
    year = {1972},
    month = {09},
    issn = {0021-9606},
    url = {https://doi.org/10.1063/1.1677750}
}

@article{Hartmann1970,
  title = {Shift and Broadening of the $^{87}\mathrm{Rb}$ 0-0 Line Due to Collisions with Krypton Buffer-Gas Atoms},
  author = {Hartmann, Francis and Hartmann-Boutron, Francoise},
  journal = {Phys. Rev. A},
  volume = {2},
  issue = {5},
  pages = {1885-1892},
  numpages = {0},
  year = {1970},
  month = {Nov},
  publisher = {American Physical Society},
  doi = {10.1103/PhysRevA.2.1885},
  url = {https://link.aps.org/doi/10.1103/PhysRevA.2.1885}
}

@article{Higginbottom2012,
  title = {Spatial-mode storage in a gradient-echo memory},
  author = {Higginbottom, D. B. and Sparkes, B. M. and Rancic, M. and Pinel, O. and Hosseini, M. and Lam, P. K. and Buchler, B. C.},
  journal = {Phys. Rev. A},
  volume = {86},
  issue = {2},
  pages = {023801},
  numpages = {10},
  year = {2012},
  month = {Aug},
  publisher = {American Physical Society},
  doi = {10.1103/PhysRevA.86.023801},
  url = {https://link.aps.org/doi/10.1103/PhysRevA.86.023801}
}

@article{Franzen1959,
  title = {Spin Relaxation of Optically Aligned Rubidium Vapor},
  author = {Franzen, W.},
  journal = {Phys. Rev.},
  volume = {115},
  issue = {4},
  pages = {850--856},
  numpages = {0},
  year = {1959},
  month = {Aug},
  publisher = {American Physical Society},
  doi = {10.1103/PhysRev.115.850},
  url = {https://link.aps.org/doi/10.1103/PhysRev.115.850}
}

@article{Franz1976,
  title = {Spin relaxation of rubidium atoms in sudden and quasimolecular collisions with light-noble-gas atoms},
  author = {Franz, F. A. and Volk, C.},
  journal = {Phys. Rev. A},
  volume = {14},
  issue = {5},
  pages = {1711-1728},
  numpages = {0},
  year = {1976},
  month = {Nov},
  publisher = {American Physical Society},
  doi = {10.1103/PhysRevA.14.1711},
  url = {https://link.aps.org/doi/10.1103/PhysRevA.14.1711}
}

@article{Hamel1986,
doi = {10.1088/0022-3700/19/24/014},
url = {https://dx.doi.org/10.1088/0022-3700/19/24/014},
year = {1986},
month = {dec},
publisher = {},
volume = {19},
number = {24},
pages = {4127},
author = {W A Hamel and  J E M Haverkort and  H G C Werij and  J P Woerdman},
title = {Calculation of alkali-noble gas diffusion cross sections relevant to light-induced drift},
journal = {J. Phys. B}
}

@article{Myneni2023,
	author = {{Allard, N. F.} and {Myneni, K.} and {Blakely, J. N.} and {Guillon, G.}},
	title = {Temperature and density dependence of line profiles of sodium perturbed by helium},
	DOI= "10.1051/0004-6361/202346215",
	url= "https://doi.org/10.1051/0004-6361/202346215",
	journal = {A and A},
	year = 2023,
	volume = 674,
	pages = "A171",
}

@Article{Medvedev2018,
author ="Medvedev, Alexander A. and Meshkov, Vladimir V. and Stolyarov, Andrey V. and Heaven, Michael C.",
title  ="Ab initio interatomic potentials and transport properties of alkali metal ({M} = {R}b and {C}s)–rare gas ({R}g = {H}e, {N}e, {A}r, {K}r, and {X}e) media",
journal  ="Phys. Chem. Chem. Phys.",
year  ="2018",
volume  ="20",
issue  ="40",
pages  ="25974-25982",
publisher  ="The Royal Society of Chemistry",
url  ="http://dx.doi.org/10.1039/C8CP04397C"}

@phdthesis{myThesis,
  author       = {Pouliot, Alexander}, 
  school       = {York University},
  type =    "{PhD} Dissertation",
  year         = 2024,
}

@inproceedings{SPIEPWest,
author = {A. Pouliot and H. C. Beica and A. Carew and A. Vorozcovs and G. Carlse and A. Kumarakrishnan},
title = {{Auto-locking waveguide amplifier system for lidar and magnetometric applications}},
volume = {10514},
booktitle = {High-Power Diode Laser Technology XVI},
editor = {Mark S. Zediker},
organization = {International Society for Optics and Photonics},
publisher = {SPIE},
pages = {152 - 159},
year = {2018},
doi = {10.1117/12.2286952},
URL = {https://doi.org/10.1117/12.2286952}
}

@article{HerminaRSI,
author = {Beica,H. C.  and Pouliot,A.  and Carew,A.  and Vorozcovs,A.  and Afkhami-Jeddi,N.  and Vacheresse,T.  and Carlse,G.  and Dowling,P.  and Barron,B.  and Kumarakrishnan,A. },
title = {Characterization and applications of auto-locked vacuum-sealed diode lasers for precision metrology},
journal = {Rev. Sci. Inst.},
volume = {90},
number = {8},
pages = {085113},
year = {2019},
doi = {10.1063/1.5112760},


}

@article{CollBroad,
title = "Collisional broadening and shift of the rubidium $\mathrm{D}1$ and $\mathrm{D}2$ lines ($5^2\mathrm{S}_{1/2}\rightarrow 5^2\mathrm{P}_{1/2}, 5^2\mathrm{P}_{3/2}$) by rare gases, $\mathrm{H}_2$, $\mathrm{D}_2$, $\mathrm{N}_2$, $\mathrm{CH}_4$ and $\mathrm{CF}_4$",
journal = "J. Quant. Spectrosc. Radiat. Transfer",
volume = "57",
number = "4",
pages = "497 - 507",
year = "1997",
issn = "0022-4073",
doi = "https://doi.org/10.1016/S0022-4073(96)00147-1",
url = "http://www.sciencedirect.com/science/article/pii/S0022407396001471",
author = "Matthew D. Rotondaro and Glen P. Perram"
}

@article{BermanPRA94,
  title = {Collisional decay and revival of the grating stimulated echo},
  author = {Berman, P. R.},
  journal = {Phys. Rev. A},
  volume = {49},
  issue = {4},
  pages = {2922-2932},
  numpages = {0},
  year = {1994},
  month = {Apr},
  publisher = {American Physical Society},
  doi = {10.1103/PhysRevA.49.2922},
  url = {https://link.aps.org/doi/10.1103/PhysRevA.49.2922}
}

@phdthesis{EricksonThesis,
  author       = {Erickson, C. J.}, 
  school       = {Princeton University},
  type =    "{PhD} Dissertation",
  year         = 2000,
}

@article{bermanLaserPhys,
  title = {Magnetic Grating Free Induction Decay and Magnetic Grating Echo},
  author = {Berman, P. R. and Dubetsky, B.},
  journal = {Laser Phys.},
  volume = {4},
  issue = {5},
  pages = {1017--1029},
  year = {1994},
 url={https://doi.org/10.1007/BF01081165},
  publisher = {Interperiodica Publishing}
}

@article{NYUVapour,
  title = {Ground-state grating echoes from $\mathrm{Rb}$ vapor at room temperature},
  author = {Kumarakrishnan, A. and Shim, U. and Cahn, S. B. and Sleator, T.},
  journal = {Phys. Rev. A},
  volume = {58},
  issue = {5},
  pages = {3868-3872},
  numpages = {0},
  year = {1998},
  month = {Nov},
  publisher = {American Physical Society},
  doi = {10.1103/PhysRevA.58.3868},
  url = {https://link.aps.org/doi/10.1103/PhysRevA.58.3868}
}

@article{NYUTrap,
  title = {Magnetic grating echoes from laser-cooled atoms},
  author = {Kumarakrishnan, A. and Cahn, S. B. and Shim, U. and Sleator, T.},
  journal = {Phys. Rev. A},
  volume = {58},
  issue = {5},
  pages = {R3387-R3390},
  numpages = {0},
  year = {1998},
  month = {Nov},
  publisher = {American Physical Society},
  doi = {10.1103/PhysRevA.58.R3387},
  url = {https://link.aps.org/doi/10.1103/PhysRevA.58.R3387}
}

@Article{NYUBeam,
author="Tonyushkin, A.
and Kumarakrishnan, A.
and Turlapov, A.
and Sleator, T.",
title="Magnetic coherence gratings in a high-flux atomic beam",
journal="Eur. Phys. J. D",
year="2010",
month="May",
day="01",
volume="58",
number="1",
pages="39--46",
issn="1434-6079",
doi="10.1140/epjd/e2010-00085-8",
url="https://doi.org/10.1140/epjd/e2010-00085-8"
}

@article{Iain2008,
  title = {Properties of magnetic sublevel coherences for precision measurements},
  author = {Chan, I. and Andreyuk, A. and Beattie, S. and Barrett, B. and Mok, C. and Weel, M. and Kumarakrishnan, A.},
  journal = {Phys. Rev. A},
  volume = {78},
  issue = {3},
  pages = {033418},
  numpages = {13},
  year = {2008},
  month = {Sep},
  publisher = {American Physical Society},
  doi = {10.1103/PhysRevA.78.033418},
  url = {https://link.aps.org/doi/10.1103/PhysRevA.78.033418}
}

@inproceedings{spiedefense,
author = {A. Pouliot and H. C. Beica and A. Carew and A. Vorozcovs and G. Carlse and B. Barrett and A. Kumarakrishnan},
title = {{Investigations of optical pumping for magnetometry using an auto-locking laser system}},
volume = {10637},
booktitle = {Laser Technology for Defense and Security XIV},
editor = {Mark Dubinskiy and Timothy C. Newell},
organization = {International Society for Optics and Photonics},
publisher = {SPIE},
pages = {40-47},
year = {2018},
doi = {10.1117/12.2304598},
URL = {https://doi.org/10.1117/12.2304598}
}

@article{Wagshul,
  title = {Laser optical pumping of high-density $\mathrm{Rb}$ in polarized $^{3}\mathrm{He}$ targets},
  author = {Wagshul, M. E. and Chupp, T. E.},
  journal = {Phys. Rev. A},
  volume = {49},
  issue = {5},
  pages = {3854-3869},
  numpages = {0},
  year = {1994},
  month = {May},
  publisher = {American Physical Society},
  doi = {10.1103/PhysRevA.49.3854},
  url = {https://link.aps.org/doi/10.1103/PhysRevA.49.3854}
}

@article{Shuker2008,
  title = {Storing Images in Warm Atomic Vapor},
  author = {Shuker, M. and Firstenberg, O. and Pugatch, R. and Ron, A. and Davidson, N.},
  journal = {Phys. Rev. Lett.},
  volume = {100},
  issue = {22},
  pages = {223601},
  numpages = {4},
  year = {2008},
  month = {Jun},
  publisher = {American Physical Society},
  doi = {10.1103/PhysRevLett.100.223601},
  url = {https://link.aps.org/doi/10.1103/PhysRevLett.100.223601}
}

@article{Holstein,
  title={Imprisonment of resonance radiation in gases. II},
  author={Holstein, T},
  journal={Physical Review},
  volume={83},
  number={6},
  pages={1159},
  year={1951},
  publisher={APS}
}

@article{Speller1979,
author={Speller, E.
and Staudenmayer, B.
and Kempter, V.},
title={Quenching cross sections for alkali-inert gas collisions},
journal={Z. Phys. A},
year={1979},
month={Dec},
day={01},
volume={291},
number={4},
pages={311-318},
issn={0939-7922},
doi={10.1007/BF01408379},
url={https://doi.org/10.1007/BF01408379}
}

@article{Ishikawa,
  title = {Diffusion coefficient and sublevel coherence of $\mathrm{Rb}$ atoms in $\mathrm{N}_{2}$ buffer gas},
  author = {Ishikawa, Kiyoshi and Yabuzaki, Tsutomu},
  journal = {Phys. Rev. A},
  volume = {62},
  issue = {6},
  pages = {065401},
  numpages = {4},
  year = {2000},
  month = {Nov},
  publisher = {American Physical Society},
  doi = {10.1103/PhysRevA.62.065401},
  url = {https://link.aps.org/doi/10.1103/PhysRevA.62.065401}
}

@article{Forber1983,
  title = {Observation of Quantum Diffractive Velocity-Changing Collisions by Use of Two-Level Heavy Optical Radiators},
  author = {Forber, R. A. and Spinelli, L. and Thomas, J. E. and Feld, M. S.},
  journal = {Phys. Rev. Lett.},
  volume = {50},
  issue = {5},
  pages = {331-335},
  numpages = {0},
  year = {1983},
  month = {Jan},
  publisher = {American Physical Society},
  doi = {10.1103/PhysRevLett.50.331},
  url = {https://link.aps.org/doi/10.1103/PhysRevLett.50.331}
}

@misc{siunit,
note={The policy of NIST is to use the International System of Units in all publications. In this document, however, units are presented in the system prevalent in the relevant discipline, although in some cases more than one system of units may be presented.}
}

@article{Mossberg1979,
  title = {Total Scattering Cross Section for {N}a on {H}e Measured by Stimulated Photon Echoes},
  author = {Mossberg, T. and Flusberg, A. and Kachru, R. and Hartmann, S. R.},
  journal = {Phys. Rev. Lett.},
  volume = {42},
  issue = {25},
  pages = {1665-1669},
  numpages = {0},
  year = {1979},
  month = {Jun},
  publisher = {American Physical Society},
  doi = {10.1103/PhysRevLett.42.1665},
  url = {https://link.aps.org/doi/10.1103/PhysRevLett.42.1665}
}

@article{Happer,
  title = {Experimental determination of the rate constants for spin exchange between optically pumped $\mathrm{K}$, $\mathrm{Rb}$, and $\mathrm{Cs}$ atoms and $^{129}\mathrm{Xe}$ nuclei in alkali-metal--noble-gas van der $\mathrm{W}$aals molecules},
  author = {Zeng, X. and Wu, Z. and Call, T. and Miron, E. and Schreiber, D. and Happer, W.},
  journal = {Phys. Rev. A},
  volume = {31},
  issue = {1},
  pages = {260-278},
  year = {1985},
  month = {Jan},
  publisher = {American Physical Society},
  doi = {10.1103/PhysRevA.31.260},
  url = {https://link.aps.org/doi/10.1103/PhysRevA.31.260}
}

@article{Franz65,
  title = {Rubidium Spin Relaxation in the Rare Gases Under Ultraclean Conditions},
  author = {Franz, F. A.},
  journal = {Phys. Rev.},
  volume = {139},
  issue = {3A},
  pages = {A603-A611},
  numpages = {0},
  year = {1965},
  month = {Aug},
  publisher = {American Physical Society},
  doi = {10.1103/PhysRev.139.A603},
  url = {https://link.aps.org/doi/10.1103/PhysRev.139.A603}
}

@article{Franz,
  title = {Analytic Expressions for Transient Signals in the Optical Pumping of Alkali-Metal Vapors},
  author = {Franz, F. A. and Sooriamoorthi, C. E.},
  journal = {Phys. Rev. A},
  volume = {8},
  issue = {5},
  pages = {2390--2401},
  numpages = {0},
  year = {1973},
  month = {Nov},
  publisher = {American Physical Society},
  doi = {10.1103/PhysRevA.8.2390},
  url = {https://link.aps.org/doi/10.1103/PhysRevA.8.2390}
}

@article{RomalisBroad,
  title = {Pressure broadening of $\mathrm{Rb}$ ${D}_{1}$ and ${D}_{2}$ lines by ${}^{3}\mathrm{He}$, ${}^{4}\mathrm{He}$, $\mathrm{N}_{2}$, and $\mathrm{Xe}$: Line cores and near wings},
  author = {Romalis, M. V. and Miron, E. and Cates, G. D.},
  journal = {Phys. Rev. A},
  volume = {56},
  issue = {6},
  pages = {4569-4578},
  numpages = {0},
  year = {1997},
  month = {Dec},
  publisher = {American Physical Society},
  doi = {10.1103/PhysRevA.56.4569},

}

@article{McNeal,
author = {McNeal,Robert J. },
title = {Disorientation Cross Sections in Optical Pumping},
journal = {J. Chem. Phys.},
volume = {37},
number = {11},
pages = {2726-2727},
year = {1962},
doi = {10.1063/1.1733089},

    
}

@article{HapperPolImg,
  title = {Spin-exchange optical pumping of noble-gas nuclei},
  author = {Walker, Thad G. and Happer, William},
  journal = {Rev. Mod. Phys.},
  volume = {69},
  issue = {2},
  pages = {629-642},
  numpages = {0},
  year = {1997},
  month = {Apr},
  publisher = {American Physical Society},
  doi = {10.1103/RevModPhys.69.629},
  url = {https://link.aps.org/doi/10.1103/RevModPhys.69.629}
}

@Article{Parniak2014,
author={Parniak, Micha{\l}
and Wasilewski, Wojciech},
title={Direct observation of atomic diffusion in warm rubidium ensembles},
journal={Appl. Phys. B},
year={2014},
month={Aug},
day={01},
volume={116},
number={2},
pages={415-421},
issn={1432-0649},
doi={10.1007/s00340-013-5712-y},
url={https://doi.org/10.1007/s00340-013-5712-y}
}

@book{Chapman, 
place={Cambridge}, 
edition={3}, 
title={The Mathematical Theory of Non-uniform Gases: An Account of the Kinetic Theory of Viscosity, Thermal Conduction and Diffusion in Gases},
publisher={Cambridge University Press},
author={Chapman, S. and Cowling, T.G. and Burnett, D.}, 
year={1990}
}

@ARTICLE{Allred,
  title = {High-Sensitivity Atomic Magnetometer Unaffected by Spin-Exchange Relaxation},
  author = {Allred, J. C. and Lyman, R. N. and Kornack, T. W. and Romalis, M. V.},
  journal = {Phys. Rev. Lett.},
  volume = {89},
  issue = {13},
  pages = {130801},
  numpages = {4},
  year = {2002},
  month = {Sep},
  publisher = {American Physical Society},
  doi = {10.1103/PhysRevLett.89.130801},
  url = {https://link.aps.org/doi/10.1103/PhysRevLett.89.130801}
}

@article{SawyerYe2008,
  title = {Molecular Beam Collisions with a Magnetically Trapped Target},
  author = {Sawyer, Brian C. and Stuhl, Benjamin K. and Wang, Dajun and Yeo, Mark and Ye, Jun},
  journal = {Phys. Rev. Lett.},
  volume = {101},
  issue = {20},
  pages = {203203},
  numpages = {4},
  year = {2008},
  month = {Nov},
  publisher = {American Physical Society},
  doi = {10.1103/PhysRevLett.101.203203},
  url = {https://link.aps.org/doi/10.1103/PhysRevLett.101.203203}
}

@article{SadeghpourSpin2009,
  title = {Collision-induced spin exchange of alkali-metal atoms with $^{3}\text{H}\text{e}$: An ab initio study},
  author = {Tscherbul, T. V. and Zhang, P. and Sadeghpour, H. R. and Dalgarno, A.},
  journal = {Phys. Rev. A},
  volume = {79},
  issue = {6},
  pages = {062707},
  numpages = {11},
  year = {2009},
  month = {Jun},
  publisher = {American Physical Society},
  doi = {10.1103/PhysRevA.79.062707},
  url = {https://link.aps.org/doi/10.1103/PhysRevA.79.062707}
}

@article{Tiesinga2017,
doi = {10.1088/1681-7575/aa8a7b},
url = {https://dx.doi.org/10.1088/1681-7575/aa8a7b},
year = {2017},
month = {nov},
publisher = {IOP Publishing},
volume = {54},
number = {6},
pages = {S125},
author = {Julia Scherschligt and James A Fedchak and Daniel S Barker and Stephen Eckel and Nikolai Klimov and Constantinos Makrides and Eite Tiesinga},
title = {Development of a new {UHV/XHV} pressure standard (cold atom vacuum standard)},
journal = {Metrologia}
}

@article{Walker2001,
  title = {Rb-{X}e spin relaxation in dilute {X}e mixtures},
  author = {Nelson, I. A. and Walker, T. G.},
  journal = {Phys. Rev. A},
  volume = {65},
  issue = {1},
  pages = {012712},
  numpages = {6},
  year = {2001},
  month = {Dec},
  publisher = {American Physical Society},
  doi = {10.1103/PhysRevA.65.012712},
  url = {https://link.aps.org/doi/10.1103/PhysRevA.65.012712}
}

@article{Saam92,
  title = {Rb${\mathrm{\ensuremath{-}}}^{129}${X}e spin-exchange rates due to binary and three-body collisions at high {X}e pressures},
  author = {Cates, G. D. and Fitzgerald, R. J. and Barton, A. S. and Bogorad, P. and Gatzke, M. and Newbury, N. R. and Saam, B.},
  journal = {Phys. Rev. A},
  volume = {45},
  issue = {7},
  pages = {4631--4639},
  numpages = {0},
  year = {1992},
  month = {Apr},
  publisher = {American Physical Society},
  doi = {10.1103/PhysRevA.45.4631},
  url = {https://link.aps.org/doi/10.1103/PhysRevA.45.4631}
}

@article{HapperSpin,
  title = {Polarization of the nuclear spins of noble-gas atoms by spin exchange with optically pumped alkali-metal atoms},
  author = {Happer, W. and Miron, E. and Schaefer, S. and Schreiber, D. and van Wijngaarden, W. A. and Zeng, X.},
  journal = {Phys. Rev. A},
  volume = {29},
  issue = {6},
  pages = {3092--3110},
  numpages = {0},
  year = {1984},
  month = {Jun},
  publisher = {American Physical Society},
  doi = {10.1103/PhysRevA.29.3092},
  url = {https://link.aps.org/doi/10.1103/PhysRevA.29.3092}
}

@article{Turlapov2016,
doi = {10.1088/0026-1394/53/6/1287},
url = {https://dx.doi.org/10.1088/0026-1394/53/6/1287},
year = {2016},
month = {nov},
publisher = {IOP Publishing},
volume = {53},
number = {6},
pages = {1287},
author = {V B Makhalov and K A Martiyanov and A V Turlapov},
title = {Primary vacuometer based on an ultracold gas in a shallow optical dipole trap},
journal = {Metrologia}
}

@article{Shen2020,
doi = {10.1088/1681-7575/ab7170},
url = {https://dx.doi.org/10.1088/1681-7575/ab7170},
year = {2020},
month = {mar},
publisher = {IOP Publishing},
volume = {57},
number = {2},
pages = {025015},
author = {Pinrui Shen and Kirk W Madison and James L Booth},
title = {Realization of a universal quantum pressure standard},
journal = {Metrologia}
}

@article{Klos,
    author = {K\l{}os, J. and Tiesinga, E.},
    title = "{Elastic and glancing-angle rate coefficients for heating of ultracold {L}i and {R}b atoms by collisions with room-temperature noble gases, {H}$_2$, and {N}$_2$}",
    journal = {J. Chem. Phys.},
    volume = {158},
    pages = {014308},
    year = {2023},
    url = {https://doi.org/10.1063/5.0124062}}

@article{Marrero1972,
    author = {Marrero, T. R. and Mason, E. A.},
    title = "{Gaseous Diffusion Coefficients}",
    journal = {J. Phys. Chem. Ref. Data},
    volume = {1},
    year = {1972},
    url = {https://doi.org/10.1063/1.3253094}}

@book{Child,
	author={M. S. Child},
	title={Molecular collision theory},
	publisher={Academic Press, London and New York},
	year=1974}

@article{CODATA2018,
  title = {{CODATA} recommended values of the fundamental physical constants: 2018},
  author = {Tiesinga, Eite and Mohr, Peter J. and Newell, David B. and Taylor, Barry N.},
  journal = {Rev. Mod. Phys.},
  volume = {93},
  pages = {025010},
  year = {2021},
  url = {https://link.aps.org/doi/10.1103/RevModPhys.93.025010}
}

@article{AME2020,
url = {https://dx.doi.org/10.1088/1674-1137/abddaf},
year = {2021},
volume = {45},
pages = {030003},
author = {Meng Wang and W.J. Huang and F.G. Kondev and G. Audi and S. Naimi},
title = {The {AME} 2020 atomic mass evaluation ({II}). Tables, graphs and references*},
journal = {Chinese Phys. C}
}

@article{Monchick1959,
    author = {Monchick, Louis},
    title = "Collision Integrals for the Exponential Repulsive Potential",
    journal = {Phys. Fluids},
    volume = {2},
    pages = {695},
    year = {1959},
    url = {https://doi.org/10.1063/1.1705974}
}

@article{DereviankoBabb2010,
title = {Electric dipole polarizabilities at imaginary frequencies for hydrogen, the alkali–metal, alkaline–earth, and noble gas atoms},
journal = {Atomic Data and Nuclear Data Tables},
volume = {96},
number = {3},
pages = {323-331},
year = {2010},
issn = {0092-640X},
doi = {https://doi.org/10.1016/j.adt.2009.12.002},
url = {https://www.sciencedirect.com/science/article/pii/S0092640X09000874},
author = {Andrei Derevianko and Sergey G. Porsev and James F. Babb}
}

@article{Gibble2013,
  title = {Scattering of Cold-Atom Coherences by Hot Atoms: Frequency Shifts from Background-Gas Collisions},
  author = {Gibble, Kurt},
  journal = {Phys. Rev. Lett.},
  volume = {110},
  issue = {18},
  pages = {180802},
  numpages = {5},
  year = {2013},
  month = {May},
  publisher = {American Physical Society},
  doi = {10.1103/PhysRevLett.110.180802},
  url = {https://link.aps.org/doi/10.1103/PhysRevLett.110.180802}
}

@article{Gibble1991,
  title = {Measurements of velocity-changing collision kernels},
  author = {Gibble, K. E. and Gallagher, A.},
  journal = {Phys. Rev. A},
  volume = {43},
  issue = {3},
  pages = {1366--1380},
  numpages = {0},
  year = {1991},
  month = {Feb},
  publisher = {American Physical Society},
  doi = {10.1103/PhysRevA.43.1366},
  url = {https://link.aps.org/doi/10.1103/PhysRevA.43.1366}
}

@article{Mlynek,
  title = {Optically driven spin nutations in the ground state of atomic sodium},
  author = {Suter, Dieter and Rosatzin, Martin and Mlynek, J\"urgen},
  journal = {Phys. Rev. A},
  volume = {41},
  issue = {3},
  pages = {1634--1644},
  numpages = {0},
  year = {1990},
  month = {Feb},
  publisher = {American Physical Society},
  doi = {10.1103/PhysRevA.41.1634},
  url = {https://link.aps.org/doi/10.1103/PhysRevA.41.1634}
}

@article{Belov1981,
  title={Application of the magnetic-scanning method to the measurement of the broadening and shift constants of the rubidium {D}2 line (780.0 nm) by foreign gases},
  author={Belov, VN},
  journal={Optics and Spectroscopy},
  volume={51},
  number={1},
  pages={22--24},
  year={1981}
}

@Manual{SRSSynth,
   title = "SG380 Series RF Signal Generators (SG384)",
   organization = "Stanford Research Systems",
   year = 2021,
   note = "",
}

@note{disclaimer,
 note="Any mention of commercial products is for information only; it does not imply recommendation or endorsement by {NIST}."
}

@article{Lengsfield:1986,
    author = {Lengsfield, Byron H., III and Yarkony, David R.},
    title = "{On the evaluation of nonadiabatic coupling matrix elements for MCSCF/CI wave functions using analytic derivative methods. III. Second derivative terms}",
    journal = {J. Chem. Phys.},
    volume = {84},
    number = {1},
    pages = {348-353},
    year = {1986},
    month = {01},
    issn = {0021-9606},
    doi = {10.1063/1.450144},
    url = {https://doi.org/10.1063/1.450144}
}

\end{document}